\documentclass[prd,twocolumn,aps,amsmath,nofootinbib,superscriptaddress, preprintnumbers]{revtex4}

\usepackage{graphicx}
\usepackage{hyperref}
\usepackage{bm}

\catcode`@=11
\let\savesort=\NAT@sort@cites
\newcommand\nosort[1]{\edef\NAT@cite@list{#1}}
\def\citenosort#1{\let\NAT@sort@cites=\nosort \cite{#1}%
   \let\NAT@sort@cites=\savesort}
\catcode`@=12 

\newcommand{\nn}{\nonumber} 
\newcommand{\bea}{\begin{eqnarray}}
\newcommand{\eea}{\end{eqnarray}}
\newcommand{\comment}[1]{}

\begin{document}


\preprint{MIT-CTP-3934} 

\title{Predicting the cosmological constant with the scale-factor
cutoff measure}

\author{Andrea De Simone}
\affiliation{Center for Theoretical Physics, Laboratory for Nuclear 
Science, and Department of Physics, \\
Massachusetts Institute of Technology, Cambridge, MA 02139}

\author{Alan H.~Guth}
\affiliation{Center for Theoretical Physics, Laboratory for Nuclear 
Science, and Department of Physics, \\
Massachusetts Institute of Technology, Cambridge, MA 02139}

\author{Michael P.~Salem}
\affiliation{Institute of Cosmology, Department of Physics and Astronomy,
Tufts University, Medford, MA 02155}

\author{Alexander Vilenkin}
\affiliation{Institute of Cosmology, Department of Physics and Astronomy,
Tufts University, Medford, MA 02155}


\begin{abstract}
It is well known that anthropic selection from a landscape with a
flat prior distribution of cosmological constant $\Lambda$ gives
a reasonable fit to observation.  However, a realistic model of
the multiverse has a physical volume that diverges with time, and
the predicted distribution of $\Lambda$ depends on how the
spacetime volume is regulated.  We study a simple model of the
multiverse with probabilities regulated by a scale-factor cutoff,
and calculate the resulting distribution, considering both
positive and negative values of $\Lambda$.  The results are in
good agreement with observation.  In particular, the scale-factor
cutoff strongly suppresses the probability for values of
$\Lambda$ that are more than about ten times the observed value.
We also discuss several qualitative features of the scale-factor
cutoff, including aspects of the distributions of the curvature
parameter $\Omega$ and the primordial density~contrast~$Q$.
\end{abstract}


\maketitle

\section{Introduction}
\label{sec:introduction}

The present understanding of inflationary cosmology suggests that
our universe is one among an infinite number of ``pockets'' in an
eternally inflating multiverse.  Each of these pockets contains
an infinite, nearly homogeneous and isotropic universe and, when
the fundamental theory admits a landscape of metastable vacua,
each may be characterized by different physical parameters, or
even different particles and interactions, than those observed
within our pocket. Predicting what physics we should expect to
observe within our region of such a multiverse is a major
challenge for theoretical physics. (For recent reviews of this
issue, see e.g.~\cite{Sergereview,Guth07,Linde07,AV06,Aguirre}.)

The attempt to build a calculus for such predictions is
complicated in part by the need to regulate the diverging
spacetime volume of the multiverse.  A number of different
approaches to this measure problem has been explored: a cutoff at
a fixed global
time~\cite{tconstflux,ccVilenkin,tconst3}\footnote{Much of the
early work sought to calculate the relative volumes occupied by
different pockets on hypersurfaces of constant
time~\cite{tconst,LLM}.  In Ref.~\cite{tconstflux} the
probabilities were expressed in terms of the fluxes appearing in
the Fokker-Planck equation for eternal inflation.  In most (but
not all) cases, this method is equivalent to imposing a cutoff at
a constant time.  The prescription of a global time cutoff was
first explicitly formulated in~\cite{ccVilenkin}.}, the so-called
``gauge-invariant'' measures~\cite{AV96,tconst4}, where different
cutoff times are used in different pockets in order to make the
measure approximately time-parametrization invariant, the
pocket-based measure~\cite{GTV,pockets,GSPVW,ELM}, which avoids
reference to global time by focusing on pocket abundances and
regulates the diverging volume within each pocket with a
spherical volume cutoff, and finally the causal patch
measures~\cite{diamond,censor}, which restrict consideration to
the spacetime volume accessible to a single observer.\footnote{We
also note the recent measure proposals in
Refs.~\cite{Vanchurin07,Winitzki08}.  Observational predictions
of these measures have not yet been worked out, so we shall not
discuss them any further.} Different measures make different
observational predictions.  In order to decide which, if any, is
on the right track, one can take an empirical approach, working
out the predictions of candidate measures and comparing them with
the data.  In this spirit, we investigate one of the simplest
global-time measure proposals:  the scale-factor cutoff measure.

The main focus of this paper is on the prediction of the
cosmological constant
$\Lambda$~\citenosort{ccweinberg,Linde84,cclinde,ccVilenkin,Efstathiou,MSW},
which is arguably a major success of the multiverse picture. 
Most calculations of the distribution of $\Lambda$ in the
literature~\cite{Efstathiou,MSW,GLV,Tegmark,VP,Peacock} do not
explicitly specify the measure, but in fact correspond to using
the pocket-based measure.  The distribution of positive $\Lambda$
in a causal-patch measure has also been
considered~\cite{ccbousso}.  The authors of Ref.~\cite{ccbousso}
emphasize that the causal-patch measure gives a strong
suppression for values of $\Lambda$ more than about ten times the
observed value, while anthropic constraints alone might easily
allow values 1000 times larger than observed, depending on
assumptions.  Here, we calculate the distribution for $\Lambda$
in the scale-factor cutoff measure, considering both positive and
negative values of $\Lambda$, and compare our results with those
of other approaches.  We find that our distribution is in a good
agreement with the observed value of $\Lambda$, and that the
scale-factor cutoff gives a suppression for large positive values
of $\Lambda$ that is very similar to that of the causal-patch
measure.

We also show that the scale-factor cutoff measure is not
afflicted with some of the serious problems arising in other
approaches. For example, another member of the global time
measure family --- the proper-time cutoff measure --- predicts a
population of observers that is extremely
youth-dominated~\citenosort{youngness1,Guth07,youngness2}.
Observers who take a little less time to evolve are hugely more
numerous than their slower-evolving counterparts, suggesting that
we should most likely have evolved at a very early cosmic time,
when the conditions for life were rather hostile.  This
counter-factual prediction is known as ``the youngness paradox''. 
Furthermore, the ``gauge-invariant'' and pocket-based measures
suffer from a ``$Q$ catastrophe,'' exponentially preferring
either very large or very small values of the primordial density
contrast $Q$~\cite{FHW,QGV}.  In fact, this problem is not
restricted to $Q$ --- there are similar expectations for the
gravitational constant $G$~\cite{GS}.  We show that the youngness
bias is very mild in the scale-factor cutoff, and that there is
no $Q$ (or $G$) catastrophe.  We also describe qualitative
expectations for the distributions of $Q$ and of the curvature
parameter $\Omega$.

This paper is organized as follows.  In section~\ref{sec:sfcutoff} 
we describe the scale-factor cutoff, commenting on its more salient 
features including its very mild youngness bias and aspects of the 
distributions of $Q$ and $\Omega$.  In section~\ref{sec:Ldist} we 
compute the probability distribution of $\Lambda$, calculating it 
first for the pocket-based measure, reproducing previous results, 
and then calculating it for the scale-factor cutoff. In both cases 
we study positive and negative values of $\Lambda$.  Our main
results are summarized in section~\ref{sec:conclusions}.  Finally, 
we include two appendices.  In appendix~\ref{sec:initial} we 
consider the possibility that the landscape splits into several 
disconnected sectors, and show that even in this situation the 
scale-factor cutoff measure is essentially independent of the 
initial state of the universe.  Appendix~\ref{sec:threshold} 
contains an analysis of the evolution of the collapse density 
threshold, along with a description of the linear growth function 
of density perturbations.

\section{The scale-factor cutoff}
\label{sec:sfcutoff}

\subsection{Global time cutoffs}
\label{ssec:global}

To introduce a global time cutoff, we start with a patch of a
spacelike hypersurface $\Sigma$ somewhere in the inflating part
of spacetime, and follow its evolution along the congruence of
geodesics orthogonal to $\Sigma$.  The spacetime region covered
by this congruence will typically have infinite spacetime volume,
and will include an infinite number of pockets.  In the
global-time cutoff approach we introduce a time coordinate $t$,
and restrict our attention to the finite spacetime region
$\Gamma(\Sigma,t_c)$ swept out by the geodesics prior to $t=t_c$,
where $t_c$ is a cutoff which is taken to infinity at the end of
the calculation.  The relative probability of any two types of 
events $A$ and $B$ is then defined to be  
\bea
{p(A) \over p(B)} \equiv \lim_{t_c \to \infty} {{n\bigl(A,
\Gamma(\Sigma, t_c)\bigr)} \over {n\bigl(B, \Gamma(\Sigma,
t_c)\bigr)}} \ ,
\eea
where $n(A,\Gamma)$ and $n(B,\Gamma)$ are the number of events of
types $A$ and $B$ respectively in the spacetime region $\Gamma$. 
In particular, the probability $P_j$ of measuring parameter
values corresponding to a pocket of type $j$ is proportional to
the number of independent measurements made in that type of
pocket, within the spacetime region $\Gamma(\Sigma,t_c)$, in the
limit $t_c \to \infty$.

The time coordinate $t$ is ``global'' in the sense that
constant-time surfaces cross many different pockets.  Note
however that it does not have to be global for the entire
spacetime, so the initial surface $\Sigma$ does not have to be a
Cauchy surface for the multiverse.  It need not be monotonic,
either, where for nonmonotonic $t$ we limit $\Gamma(\Sigma,t_c)$
to points along the geodesics prior to the first occurrence of
$t=t_c$.

As we will discuss in more detail in appendix~\ref{sec:initial},
probability distributions obtained from this kind of measure are
independent of the choice of the hypersurface 
$\Sigma$.\footnote{Here, and in most of the paper, we assume an 
irreducible
landscape, where any metastable inflating vacuum is accessible
from any other such vacuum through a sequence of transitions. 
Alternatively, if the landscape splits into several disconnected
sectors, each sector will be characterized by an independent
probability distribution and our discussion will still be
applicable to any of these sectors. The distribution in case of a
reducible landscape is discussed in appendix~\ref{sec:initial}.}
They do depend, however, on how one defines the time parameter
$t$.  To understand this sensitivity to the choice of cutoff,
note that the eternally inflating universe is rapidly expanding,
such that at any time most of the volume is in pockets that have
just formed.  These pockets are therefore very near the cutoff
surface at $t=t_c$, which explains why distributions depend on
exactly how that surface is drawn.

A natural choice of the time coordinate $t$ is the proper time
$\tau$ along the geodesic congruence.  But as we have already
mentioned, and will discuss in more detail in the following
subsection, this choice is plagued with the youngness paradox,
and therefore does not yield a satisfactory measure.  Another
natural option is to use the expansion factor $a$ along the
geodesics as a measure of time.  The scale-factor time is then
defined as
\bea
t\equiv\ln a \,.
\label{tdef}
\eea
The use of this time parameter for calculating probabilities is
advocated in Ref.~\cite{Starobinsky} and is studied in various
contexts in Refs.~\cite{tconst}, \cite{LLM},
\cite{tconstflux}, and \cite{tconst3}.\footnote{The measure
studied in Ref.~\cite{Starobinsky} is a comoving-volume measure
on surfaces of constant scale-factor time; it is different from
the scale-factor cutoff measure being discussed here.  In
particular, the former measure has a strong dependence on the
initial state at the hypersurface $\Sigma$.  Our measure is very
similar to one studied in Ref.~\cite{tconst3}, which is called
the ``pseudo-comoving volume-weighted measure.''} It amounts to
measuring time in units of the local Hubble time $H^{-1}$,
\bea 
dt=Hd\tau \,.
\label{ttau}
\eea
The scale-factor cutoff is imposed at a fixed value of $t=t_c$,
or, equivalently, at a fixed expansion factor $a_c$.

The term ``scale factor'' is often used in the context of
homogeneous and isotropic spaces, but it is easily generalized to
spacetimes with no such symmetry. In the general case, the
scale-factor time can be defined by Eq.~(\ref{ttau}) with
\bea
H=(1/3)\,u^\mu{}_{;\,\mu} \, ,
   \label{Hubble}
\eea
where $u^\mu(x)$ is the four-velocity vector along the geodesics. 
This definition has a simple geometric meaning, which can be seen
by imagining that the congruence of geodesics describes the flow
of a ``dust'' of test particles. If the dust of particles is
assumed to have a uniform density $\rho_0$ on the initial surface
$\Sigma$, then the four-current of the dust can be described by
$j^\mu(x) = \rho(x) u^\mu(x)$, where $\rho=\rho_0$ on $\Sigma$. 
Conservation of the current then implies that $u^\mu \partial_\mu
\rho + \rho \, u^\mu{}_{;\mu} = 0$, which with Eqs.~(\ref{ttau})
and (\ref{Hubble}) implies that
\bea
D_\tau \ln \rho = - u^\mu{}_{;\mu} = - 3 D_\tau t \, ,
\eea
where $D_\tau \equiv u^\mu \partial_\mu$ is the derivative with
respect to proper time along the geodesics.  The solution is then
$\rho = \rho_0 e^{- 3 t}$.  From Eq.~(\ref{tdef}) we then have $a
\propto
\rho^{-1/3}$, so the scale-factor cutoff is triggered when the
density $\rho(x)$ of the dust in its own rest frame drops below a
certain specified level.

The divergence of geodesics during inflation or homogeneous
expansion can be followed by convergence during structure
formation or in regions dominated by a negative cosmological
constant.  The scale-factor time then ceases to be a good time
variable, but this does not preclude one from using it to impose
a cutoff.  A geodesic is terminated when the scale factor first
reaches the cutoff value $a_c$.  If the scale factor turns around
and starts decreasing before reaching that value, we continue the
geodesic all the way to the crunch. When geodesics cross we can 
still define the scale factor time along each geodesic according 
to Eqs.~(\ref{ttau}) and (\ref{Hubble}); then one includes a point 
in $\Gamma(\Sigma,t_c)$ if it lies on any geodesic prior to the 
first occurrence of $t=t_c$ on that geodesic.

To facilitate further discussion, it will be useful to review
some general features of eternally inflating spacetimes, and how
they are reflected in proper time and scale-factor time slicings. 
Regions of an eternally inflating multiverse may evolve in two
distinct ways.  In the case of quantum
diffusion~\cite{eternal1,Linde86}, inflation is driven by the
potential energy of some light scalar fields, the evolution of
which is dominated by quantum fluctuations and is described by
the Fokker-Planck equation (see e.g. Ref.~\cite{LLM}). Pockets
form when the scalar field(s) fluctuate into a region of
parameter space where classical evolution dominates, and
slow-roll inflation ensues.  One can define spacelike
hypersurfaces separating the quantum and classical regimes (see
for example Ref.~\cite{GSPVW}), which we denote by $\Sigma_q$. 
In universes like ours, slow-roll inflation is followed by
thermalization (reheating) and the standard post-inflationary
evolution.  We denote the hypersurface of thermalization, which
separates the inflationary and post-inflationary epochs, as
$\Sigma_*$.

The multiverse may also (or instead) feature massive fields
associated with large false-vacuum energies.  Evolution is then
governed by bubble nucleation through quantum
tunneling~\cite{Gott,Steinhardt} and can be described to good
approximation by a suitable master equation \citenosort{GV97,GSPVW}.  
The tunneling may proceed into another local minimum, into a region
of quantum diffusion, or into a region of classical slow-roll
inflation. In the latter case, the bubble interiors have the
geometry of open FRW universes \cite{ColemanDeLuccia}. Bubbles of
interest to us here have a period of slow-roll inflation followed
by thermalization.  The role of the hypersurface $\Sigma_q$ is
played in this case by the surface separating the initial
curvature-dominated regime and the slow-roll regime inside the
bubble. The differences between quantum diffusion and tunneling
are not important for most of the discussion below, so we shall
use notation and terminology interchangeably.

The number of objects of any type that have formed prior to some
time $t$ is proportional to $e^{\gamma t}$, where $\gamma$ is the
largest eigenvalue of the physical-volume Fokker-Planck or master
equation. This is because the asymptotic behavior is determined
by the eigenstate with the largest eigenvalue.  Similarly, the
physical volume that thermalizes into pockets of type $j$ between
times $t$ and $t+dt$ has the form
\bea
dV_{*j} = C_j e^{\gamma t}dt \,,
\label{dV}
\eea
where $C_j$ is a constant that depends on the type of pocket.
(This was derived in Ref.~\cite{tconstflux} for models with
quantum diffusion and in Refs.~\cite{VW} and \cite{youngness2}
for models with bubble nucleation.)

The value of $\gamma$ in Eq.~(\ref{dV}) is the same for all
pockets, but it depends on the choice of time variable $t$.  With
a proper-time slicing, it is given by
\bea
\gamma\sim 3 H_{\rm max} \qquad (t=\tau)\,,
\label{gammaproper}
\eea
where $H_{\rm max}$ is the expansion rate of the highest-energy
vacuum in the landscape, and corrections associated with decay 
rates and upward tunneling rates have been ignored.  In this
case the overall expansion of the multiverse is driven by this
fastest-expanding vacuum, which then ``trickles down'' to all of
the other vacua.  With scale-factor slicing, all regions would
expand as $a^3=e^{3t}$ if it were not for the continuous loss of
volume to terminal vacua with negative or zero $\Lambda$. 
Because of this loss, the value of $\gamma$ is slightly smaller
than 3, and the difference is determined mostly by the rate of
decay of the slowest-decaying (dominant) vacuum in the
landscape~\cite{Delia},
\bea
\gamma \approx 3-\kappa_D \qquad (t=\ln a) \,.
\label{gammascale}
\eea
Here,
\bea
\kappa_D=(4\pi/3)\,\Gamma_D/H_D^4 \,,
\label{kappa}
\eea
where $\Gamma_D$ is the decay rate of the dominant vacuum per
unit spacetime volume, and $H_D$ is its expansion rate. The
vacuum decay rate is typically exponentially suppressed, so for
the slowest-decaying vacuum we expect it to be extremely small. 
Hence,
\bea
3-\gamma \ll 1 \,.
\label{gammasmall}
\eea

\subsection{The youngness bias} 
\label{ssec:youngness}

As we have already mentioned, the proper-time cutoff measure
leads to rather bizarre predictions, collectively known as the
youngness paradox~\citenosort{youngness1,Guth07,youngness2}.  With
proper time slicing, Eqs.~(\ref{dV}) and (\ref{gammaproper}) tell
us that the growth of volume in regions of all types is extremely
fast, so at any time the thermalized volume is exponentially
dominated by regions that have just thermalized.  With this
super-fast expansion, observers who take a little less time to
evolve are rewarded by a huge volume factor. This means most
observers form closer to the cutoff, when there is much more
volume available.  Assuming that $H_{\rm max}$ is comparable to
Planck scale, as one might expect in the string theory landscape,
then observers who evolved faster than us by $\Delta\tau=10^9$
years would have an available thermalized volume which is larger
than the volume available to us by a factor of
\bea
e^{\gamma \, \Delta \tau} \sim e^{3 H_{\rm max} \, \Delta \tau}
\sim \exp(10^{60}) \,.
\eea
Unless the probability of life evolving so fast is suppressed by
a factor greater than $\exp(10^{60})$, then these rapidly
evolving observers would outnumber us by a huge factor.  Since
these observers would measure the cosmic microwave background (CMB) 
temperature to be $T=2.9$~K, it would be hard to explain why we 
measure it to be $T=2.73$~K.  Note that because 
$H_{\rm max}\Delta\tau$ appears in the exponent, the situation is 
qualitatively unchanged by considering much smaller values of 
$H_{\rm max}$ or $\Delta\tau$. 

The situation with a scale-factor cutoff is very different.  To
illustrate methods used throughout this paper, let us be more
precise. Let $\Delta t$ denote the interval in scale-factor time
between the time of thermalization, $t_*$, and the time when some
class of observers measures the CMB temperature.  A time cutoff
excludes the counting of observers who measure the CMB
temperature at times later than $t_c$, so the number of counted
observers is proportional to the volume that thermalizes at time
$t_* < t_c - \Delta t$. (For simplicity we focus on pockets that
have the same low-energy physics as ours.) The volume of regions
thermalized per unit time is given by Eq.~(\ref{dV}).  During the
time interval $\Delta t$, some of this volume may decay by
tunneling transitions to other vacua.  This effect is negligible, 
and we henceforth ignore it.  For a given $\Delta t$, the 
thermalized volume available for observers to evolve, as counted 
by the scale-factor cutoff measure, is
\bea
{\mathcal V}(\Delta t) \,\,\propto\, 
\int_{-\infty}^{t_c-\Delta t}\! e^{\gamma t_*}\, dt_* 
\,\,\propto\,\,\, e^{-\gamma \Delta t} \,.
\eea

To compare with the results above, consider the relative amounts
of volume available for the evolution of two different 
civilizations, which form at two different time intervals since
thermalization, $\Delta t_1$ and $\Delta t_2$:
\bea
\frac{{\mathcal V}(\Delta t_1)}{{\mathcal V}(\Delta t_2)} \,\,=\,\, 
e^{\gamma(\Delta t_2-\Delta t_1)}
\,\,=\,\, \left(a_2/a_1\right)^{\gamma} \,,
\eea
where $a_i$ is the scale factor at time $t_*+\Delta t_i$.  Thus,
taking $\gamma \approx 3$, the relative volumes available for 
observers who measure the CMB at the present value ($T=2.73$~K), 
compared to observers who measure it at the value of $10^9$ years 
ago ($T=2.9$~K), is given by
\bea
\frac{{\mathcal V}(2.73~\hbox{K})}{{\mathcal V}(2.9~\hbox{K})} \approx
\left({2.73~\hbox{K}\over{2.9~\hbox{K}}}\right)^{\!3}\approx\, 0.8 \,.
\eea
Thus, the youngness bias is very mild in the scale-factor cutoff
measure.  Yet, as we shall see, it can have interesting
observational implications.

\subsection{Expectations for the density contrast $Q$ and the
curvature parameter $\Omega$}
\label{ssec:runaway}

Pocket-based measures, as well as ``gauge-invariant'' measures,
suffer from a ``$Q$ catastrophe'' where one expects to measure
extreme values of the primordial density contrast $Q$.  To see
this, note that these measures exponentially prefer parameter
values that generate a large number of e-folds of inflation. 
This by itself does not appear to be a problem, but $Q$ is
related to parameters that determine the number of e-folds.  The
result of this is a selection effect that exponentially prefers
the observation of either very large or very small values of $Q$,
depending on the model of inflation and on which inflationary
parameters scan (i.e., which parameters vary significantly across 
the landscape)~\cite{FHW,QGV}. On the other hand, we observe $Q$ 
to lie comfortably in the middle of the anthropic
range~\cite{anthQ}, indicating that no such strong selection
effect is at work.\footnote{Possible resolutions to this problem
have been proposed in Refs.~\cite{FHW,QGV,HWY,LM}.} Note that a
similar story applies to the magnitude of the gravitational
constant $G$~\cite{GS}. 

With the scale-factor cutoff, on the other hand, this is not a
problem. To see this, consider a landscape in which the only
parameter that scans is the number of e-folds of inflation; all
low-energy physics is exactly as in our universe.  Consider first
the portions of the hypersurfaces $\Sigma_q$ that begin slow-roll
inflation at time $t_q$ in the interval $dt_q$.  These regions
begin with a physical volume proportional to $e^{\gamma
t_q}\,dt_q$, and those that do not decay grow by a factor of
$e^{3N_e}$ before they thermalize at time $t_*=t_q+N_e$.  If
$\kappa_I$ is the transition rate out of the slow-roll
inflationary phase (as defined in Eq.~(\ref{kappa})), then the
fraction of volume that does not undergo decay is
$e^{-\kappa_IN_e}$.

After thermalization at time $t_*$, the evolution is the same in
all thermalized regions.  Therefore we ignore this common
evolution and consider the number of observers measuring a given
value of $N_e$ to be proportional to the volume of thermalization
hypersurfaces that appear at times earlier than the cutoff at
scale-factor time $t_c$.  This cutoff requires $t_*=t_q+N_e<t_c$. 
Summing over all times $t_q$ gives
\bea
P(N_e) \,\propto\, e^{(3-\kappa_I)N_e}\!\!
\int_{-\infty}^{t_c-N_e}\!\!  
e^{\gamma t_q} \, dt_q \,\propto\, e^{(3-\gamma-\kappa_I)N_e} \,.
\,\,
\eea
Even though the dependence on $N_e$ is exponential, the factor
\bea
3-\gamma-\kappa_I\approx \kappa_D-\kappa_I
\eea 
is exponentially suppressed.  Thus we find $P(N_e)$ is a very
weak function of $N_e$, and there is not a strong selection
effect for a large number of e-folds of slow-roll inflation. In
fact, since the dominant vacuum $D$ is by definition the
slowest-decaying vacuum, we have $\kappa_I > \kappa_D$.  Thus the
scale-factor cutoff introduces a very weak selection for smaller
values of $N_e$.\footnote{We are grateful to Ben Freivogel for
pointing out to us the need to account for vacuum decay during
slow-roll inflation.  He has also suggested that this effect will
lead to preference for smaller values of $N_e$.}

Because of the very mild dependence on $N_e$, we do not expect
the scale-factor measure to impose significant cosmological
selection on the scanning of any inflationary parameters.  Thus,
there is no $Q$ catastrophe --- nor is there the related problem
for $G$ --- and the distribution of $Q$ is essentially its
distribution over the states in the landscape, modulated by
inflationary dynamics and any anthropic selection effects.

The distribution $P(N_e)$ is also important for the expected
value of the curvature parameter $\Omega$.  This is because the
deviation of $\Omega$ from unity decreases during an inflationary
era,
\bea
|\Omega -1|\propto e^{-2N_e} \,.
\eea
Hence pocket-based and ``gauge-invariant'' measures, which
exponentially favor large values of $N_e$, predict a universe
with $\Omega$ extremely close to unity.  The distributions of
$\Omega$ from a variety of models have been calculated using a
pocket-based measure in Refs.~\cite{GTV} and \cite{VW}.

On the other hand, as we have just described, the scale-factor
cutoff measure does not significantly select for any value of
$N_e$.  There will still be some prior distribution of $N_e$,
related to the distributions of inflationary parameters over the
states in the landscape, but it is not necessary that $N_e$ be
driven strongly toward large values (in fact, it has been argued
that small values should be preferred in the string landscape,
see e.g. Ref.~\cite{FKMS}). Thus, it appears that the
scale-factor cutoff allows for the possibility of a detectable
negative curvature.  The probability distribution of $\Omega$ in
this type of measure has been discussed qualitatively in
Ref.~\cite{FKMS}; a more detailed quantitative analysis will be
given elsewhere~\cite{Omeganext}.

\boldmath
\section{The Distribution of $\Lambda$}
\label{sec:Ldist}
\unboldmath

\boldmath
\subsection{Model assumptions}      
\label{ssec:model}
\unboldmath

We now consider a landscape of vacua with the same low-energy
physics as we observe, except for an essentially continuous
distribution of possible values of $\Lambda$.  According to 
Eq.~(\ref{dV}), the volume that thermalizes between times $t_*$ 
and $t_*+dt_*$ with values of cosmological constant between 
$\Lambda$ and $\Lambda+d\Lambda$ is given by
\bea
dV_*(\Lambda) = C(\Lambda)d\Lambda\, e^{\gamma t_*}dt_* \,.
\eea
The factor of $C(\Lambda)$ plays the role of the ``prior'' 
distribution of $\Lambda$; it depends on the spectrum of 
possible values of $\Lambda$ in the landscape and on the 
dynamics of eternal inflation.  The standard 
argument~\cite{ccweinberg,Efstathiou} suggests that 
$C(\Lambda)$ is well approximated by
\bea
C(\Lambda)\approx {\rm const} \,,
\label{flatprior}
\eea
because anthropic selection restricts $\Lambda$ to values that
are very small compared to its expected range of variation in 
the landscape.  The conditions of validity of this heuristic
argument have been studied in simple landscape 
models~\cite{Delia,KenDelia,Shenker}, with the conclusion that 
it does in fact apply to a wide class of models.  Here, we shall
assume that Eq.~(\ref{flatprior}) is valid.

Anthropic selection effects are usually characterized by the 
fraction of matter that has clustered in galaxies.  The idea here 
is that a certain average number of stars is formed per unit 
galactic mass and a certain number of observers per star, and that 
these numbers are not strongly affected by the value of $\Lambda$. 
Furthermore, the standard approach is to assume that some minimum
halo mass $M_G$ is necessary to drive efficient star formation
and heavy element retention.  Since we regulate the volume of the
multiverse using a time cutoff, it is important for us to also
track at what time observers arise.  We assume that after halo
collapse, some fixed proper time lapse $\Delta\tau$ is required
to allow for stellar, planetary, and biological evolution before
an observer can measure $\Lambda$.  Then the number of observers
measuring $\Lambda$ before some time $\tau$ in a thermalized
volume of size $V_*$ is roughly
\bea
{\mathcal N} \propto F(M_G,\tau-\Delta\tau) V_* \,,
\label{Nobs1}
\eea  
where $F$ is the collapse fraction, measuring the fraction of
matter that clusters into objects of mass greater than or equal
to $M_G$, at time $\tau-\Delta\tau$.

Anthropic selection for structure formation ensures that within
each relevant pocket matter dominates the energy density before
$\Lambda$ does.  Thus, all thermalized regions evolve in the same
way until well into the era of matter domination.  To draw upon
this common evolution, within each pocket we define proper time
$\tau$ with respect to a fixed time of thermalization, $\tau_*$. 
It is convenient to also define a reference time $\tau_m$ such
that $\tau_m$ is much larger than the time of matter-radiation
equality and much less than the time of matter-$\Lambda$
equality.  Then evolution before time $\tau_m$ is the same in
every pocket, while after $\tau_m$ the scale factor evolves as
\bea
\tilde{a}(\tau)=\left\{ 
\begin{array}{l l}
H_\Lambda^{-2/3}\sinh^{2/3}\!
\big(\textstyle{\frac{3}{2}}H_\Lambda\tau\big)\quad 
& {\rm for }\,\, \Lambda > 0 \\ H_\Lambda^{-2/3}\sin^{2/3}\!
\big(\textstyle{\frac{3}{2}}H_\Lambda\tau\big)\quad
& {\rm for }\,\, \Lambda < 0 \,.
\end{array}
\right.
\label{a}
\eea
Here we have defined
\bea
H_\Lambda\equiv \sqrt{|\Lambda|/3} \,,
\eea
and use units with $G=c=1$.  The prefactors $H_\Lambda^{-2/3}$
ensure that early evolution is identical in all thermalized
regions.  This means the global scale factor $a$ is related to
$\tilde{a}$ by some factor that depends on the scale-factor time
$t_*$ at which the region of interest thermalized.

In the case $\Lambda >0$, the rate at which halos accrete matter
decreases with time and halos may settle into galaxies that
permit quiescent stellar systems such as ours.  The situation
with $\Lambda < 0$ is quite different.  At early times, the
evolution of overdensities is the same; but when the proper time
reaches $\tau_{\rm turn}=\pi/3H_\Lambda$, the scale factor begins
to decrease and halos begin to accrete matter at a rate that
increases with time.  Such rapid accretion may prevent galaxies
from settling into stable configurations, which in turn would
cause planetary systems to undergo more frequent close encounters
with passing stars.  This effect might become significant even 
before turnaround, since our present environment benefits from 
positive $\Lambda$ slowing the collision rate of the Milky Way 
with other systems.

For this reason, we use Eq.~(\ref{Nobs1}) to estimate the number
of observers if $\Lambda > 0$, but for $\Lambda < 0$ we consider
two alternative anthropic hypotheses:
\begin{itemize}
\item[{$A$.\quad}] we use Eq.~(\ref{Nobs1}), but of course taking
account of the fact that the proper time $\tau$ cannot exceed
$\tau_{\rm crunch}=2\pi/3H_\Lambda$; or
\item[{$B$.\quad}] we use Eq.~(\ref{Nobs1}), but with the
hypothesis that the proper time $\tau$ is capped at $\tau_{\rm
turn} = \pi/3 H_\Lambda$.
\end{itemize}
Here $\tau_{\rm crunch}$ refers to the proper time at which a
thermalized region in a collapsing pocket reaches its future
singularity, which we refer to as its ``crunch.'' Anthropic
hypothesis $A$ corresponds to the assumption that life can form in
any sufficiently massive collapsed halo, while anthropic
hypothesis $B$ reflects the assumption that the probability for
the formation of life becomes negligible in the tumultuous
environment following turnaround. Similar hypotheses for
$\Lambda<0$ were previously used in Ref.~\cite{Peacock}. It seems
reasonable to believe that the truth lies somewhere between these
two hypotheses, perhaps somewhat closer to hypothesis B.

\boldmath
\subsection{Distribution of $\Lambda$ using a pocket-based measure}      
\label{ssec:standardL}
\unboldmath

Before calculating the distribution of $\Lambda$ using a
scale-factor cutoff, we review the standard
calculation~\cite{Efstathiou,MSW,GLV,Tegmark,VP,Peacock}.  This
approach assumes an ensemble of equal-size regions with a flat
prior distribution of $\Lambda$.  The regions are allowed to
evolve indefinitely, without any time cutoff, so in the case of
$\Lambda >0$ the selection factor is given by the asymptotic
collapse fraction at $\tau\to\infty$.  For $\Lambda <0$ we shall
consider anthropic hypotheses $A$ and $B$.  This prescription
corresponds to using the pocket-based measure, in which the
ensemble includes spherical regions belonging to different
pockets and observations are counted in the entire comoving
history of these regions.  The corresponding distribution
function is given by
\bea
P(\Lambda) \propto \left\{
\begin{array}{ll}
F(M_G,\tau\to\infty) \quad & {\rm for}\,\, \Lambda> 0 \\
F(M_G,\tau_{\rm crunch}-\Delta\tau) \quad & {\rm for}\,\,
\Lambda< 0 \,\,\, (A) \\ F(M_G,\tau_{\rm turn}-\Delta\tau) \quad
& {\rm for}\,\, \Lambda< 0 \,\,\, (B) \,, \\
\end{array}\right.
\label{PocketBased}
\eea     
where, again, $\tau_{\rm crunch}=2\pi/3H_\Lambda$ is the proper
time of the crunch in pockets with $\Lambda<0$, while $\tau_{\rm
turn}=\pi/3H_\Lambda$.

We approximate the collapse fraction $F$ using the
Press-Schechter (PS) formalism~\cite{PS}, which gives
\bea
F(M_G,\tau) = {\rm erfc}\left[
\frac{\delta_c(\tau)}{\sqrt{2}\,\sigma(M_G,\tau)}\right] \,,
\label{warren}
\eea     
where $\sigma(M_G,\tau)$ is the root-mean-square fractional
density contrast $\delta M/M$ averaged over a comoving scale
enclosing mass $M_G$ and evaluated at proper time $\tau$, while
$\delta_c$ is the collapse density threshold. As is further
explained in appendix~\ref{sec:threshold}, $\delta_c(\tau)$ is
determined by considering a ``top-hat'' density perturbation in a
flat universe, with an arbitrary initial amplitude. 
$\delta_c(\tau)$ is then defined as the amplitude reached by the
linear evolution of an overdensity of nonrelativistic matter
$\delta \rho_m/\rho_m$ that has the same initial amplitude as a
top-hat density perturbation that collapses to a singularity in
proper time $\tau$.  $\delta_c(\tau)$ has the constant value of
1.686 in an Einstein-de Sitter universe (i.e., flat,
matter-dominated universe), but it evolves with time when
$\Lambda\neq 0$~\cite{dc,PTV}.  We simulate this evolution using
the fitting functions (\ref{fitdeltac}), which are accurate to
better than 0.2\%. Note, however, that the results are not
significantly different if one simply uses the constant value
$\delta_c=1.686$.

\begin{figure}[t!]
\includegraphics[width=0.4\textwidth]{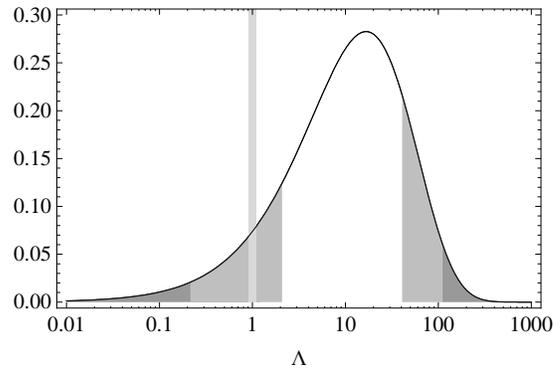} 
\caption{The normalized distribution of $\Lambda$ for $\Lambda>0$, 
with $\Lambda$ in units of the observed value, for the pocket-based
measure. The vertical bar highlights the value we measure, while
the shaded regions correspond to points more than one and two 
standard deviations from the mean.}
\label{fig:ccpocketL>0}
\end{figure}

\begin{figure*}[t!]
\begin{tabular}{ccc}
\vspace{5pt} anthropic hypothesis $A$ & & anthropic hypothesis $B$ \\
\vspace{-5pt}
\!\!\!\!\!\includegraphics[width=0.39\textwidth]{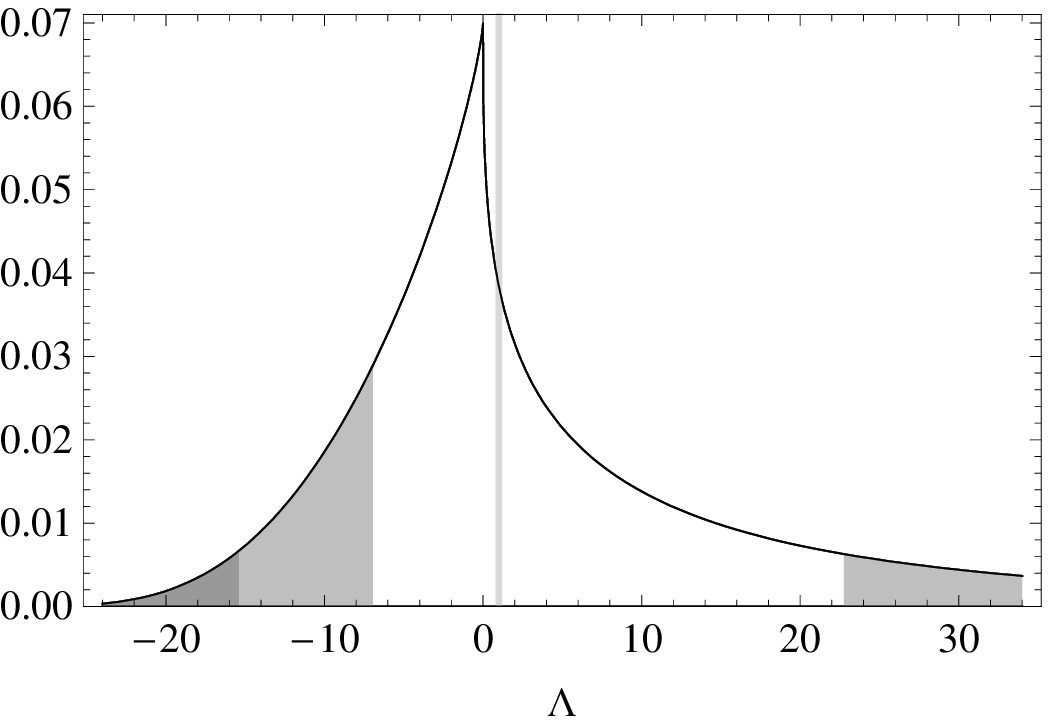} & 
\hspace{1cm} &
\includegraphics[width=0.4\textwidth]{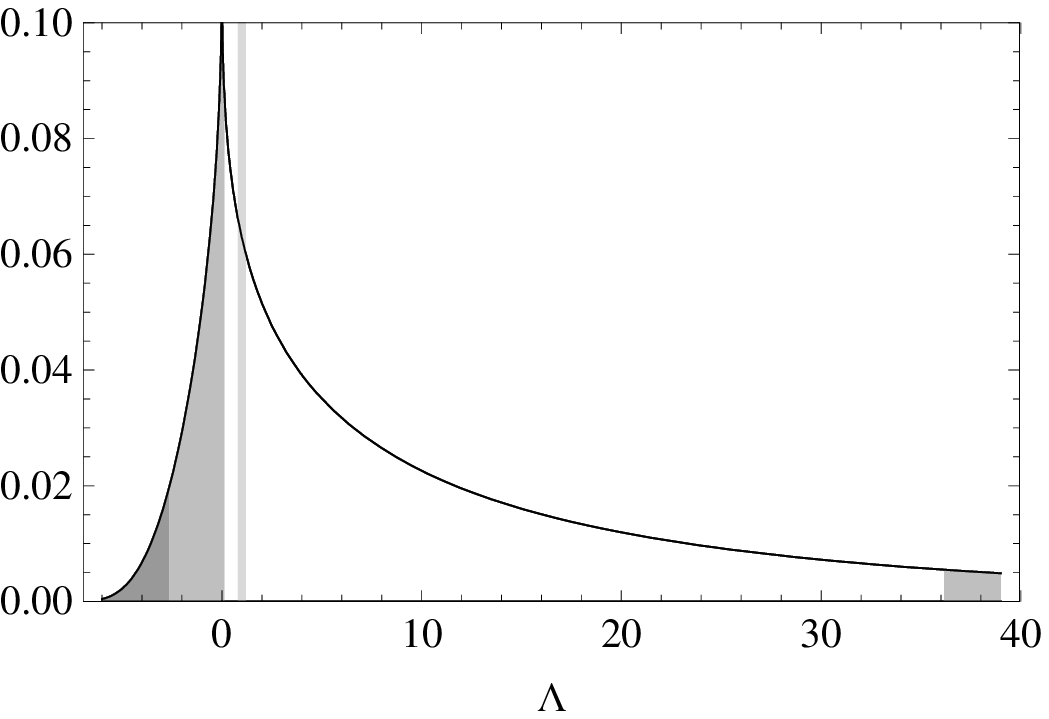} \\
\includegraphics[width=0.4\textwidth]{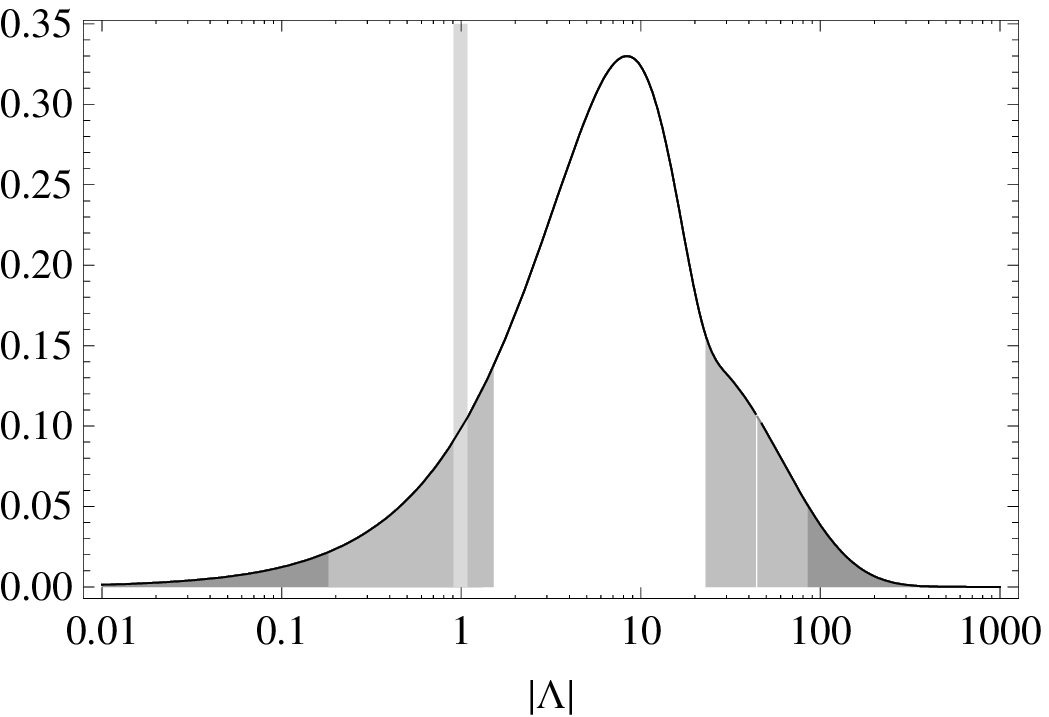} & &
\includegraphics[width=0.4\textwidth]{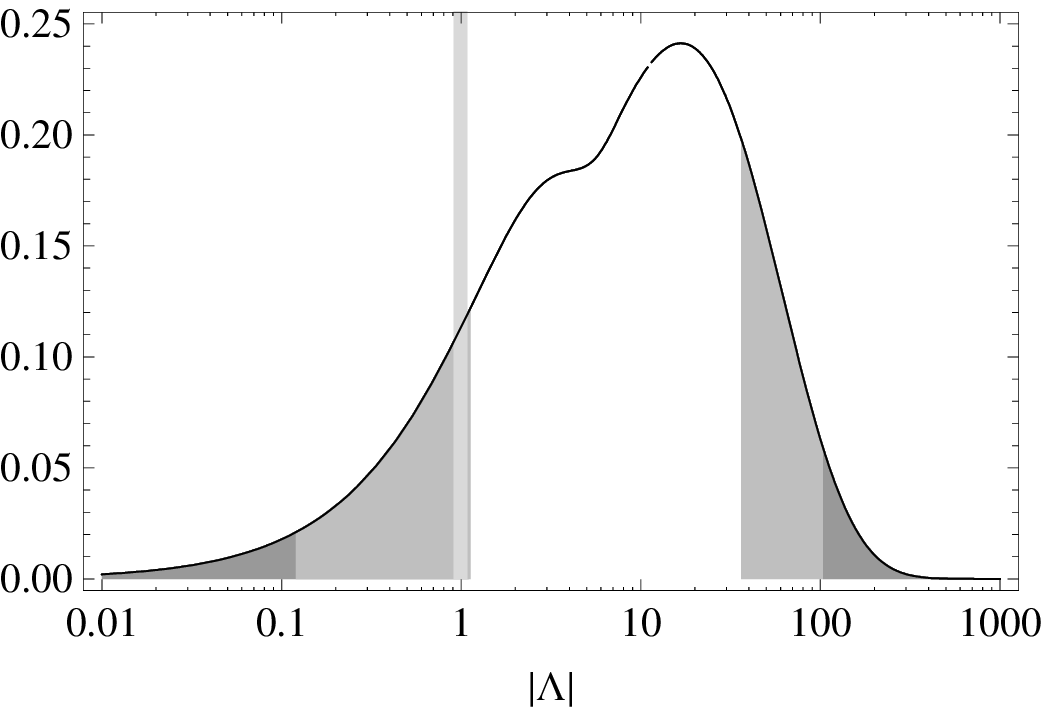} \\
\end{tabular}
\caption{The normalized distribution of $\Lambda$, with $\Lambda$ 
in units of the observed value, for the pocket-based measure.  
The left column corresponds to anthropic hypothesis $A$ while the
right column corresponds to anthropic hypothesis $B$.  Meanwhile, 
the top row shows $P(\Lambda)$ while the bottom row shows 
$P(|\Lambda|)$. The vertical bars highlight the value we
measure, while the shaded regions correspond to points more than
one and two standard deviations from the mean.}
\label{fig:ccpocket}
\end{figure*}

Aside from providing the collapse fraction, the PS formalism
describes the ``mass function,'' i.e. the distribution of halo
masses as a function of time.  $N$-body simulations indicate that
PS model overestimates the abundance of halos near the peak of
the mass function, while underestimating that of more massive
structures~\cite{Jenkins}.  Consequently, other models have been
developed (see e.g. Refs.~\cite{ST}), while others have studied
numerical fits to $N$-body results~\cite{Warren,Peacock}.  From
each of these approaches, the collapse fraction can be obtained
by integrating the mass function.  We have checked that our
results are not significantly different if we use the fitting
formula of Ref.~\cite{Peacock} instead of Eq.~(\ref{warren}). 
Meanwhile, we prefer Eq.~(\ref{warren}) to the fit of
Ref.~\cite{Peacock} because the latter was performed using only
numerical simulations with $\Lambda>0$. 

The evolution of the density contrast $\sigma$ is treated
linearly, to be consistent with the definition of the collapse
density threshold $\delta_c$.  Thus we can factorize the behavior
of $\sigma(M_G,\tau)$, writing
\bea
\sigma(M_G,\tau)=\bar\sigma(M_G)\,G_\Lambda(\tau)\,, 
\label{factorize}
\eea
where $G_\Lambda(\tau)$ is the linear growth function, which is
normalized so that the behavior for small $\tau$ is given by
$G_\Lambda(\tau)\sim (3 H_\Lambda \tau/2)^{2/3}$.  In 
appendix~\ref{sec:threshold}
we will give exact integral expressions for $G_\Lambda(\tau)$,
and also the fitting formulae~(\ref{sigp}) and~(\ref{sign}),
taken from Ref.~\cite{Peacock}, that we actually used in our
calculations.  Note that for $\Lambda\geq 0$ the growth rate
$\dot G_\Lambda(\tau)$ always decreases with time ($\ddot
G_\Lambda(\tau) < 0$), while for $\Lambda < 0$ the growth rate
reaches a minimum at $\tau \approx 0.24 \tau_{\rm crunch}$ and
then starts to accelerate. This accelerating rate of growth is
related to the increasing rate of matter accretion in collapsed
halos after turnaround, which we mentioned above in motivating
the anthropic hypothesis $B$.

The prefactor $\bar\sigma(M_G)$ in Eq.~(\ref{factorize}) depends
on the scale $M_G$ at which the density contrast is evaluated.
According to our anthropic model, $M_G$ should correspond to the
minimum halo mass for which star formation and heavy element
retention is efficient.  Indeed, the efficiency of star formation
is seen to show a sharp transition: it falls abruptly for halo
masses smaller than $M_G\sim 2\times 10^{11}M_\odot$, where
$M_\odot$ is the solar mass~\cite{StarForm}.  Peacock
\cite{Peacock} showed that the existing data on the evolving
stellar density can be well described by a Press-Schechter
calculation of the collapsed density for a single mass scale,
with a best fit corresponding to $\sigma(M_G,\,
\tau_{1000})\approx 6.74\times 10^{-3}$, where $\tau_{1000}$ is
the proper time corresponding to a temperature $T=1000$ K.  Using
cosmological parameters current at the time, Peacock found that
this perturbation amplitude corresponds to an effective galaxy
mass of $1.9 \times 10^{12}\,M_\odot$.  Using the more recent 
WMAP-5 parameters~\cite{WMAP5}, as is done throughout 
this paper,\footnote{The relevant values are
$\Omega_\Lambda=0.742$, $\Omega_m=0.258$, $\Omega_b=0.044$,
$n_s=0.96$, $h=0.719$, and $\Delta_{\mathcal R}^2(k=0.02\,{\rm
Mpc}^{-1})= 2.21\times 10^{-9}$.} 
we find (using Ref.~\cite{cmbfast} and the CMBFAST program) that 
the corresponding effective galaxy mass is $1.8\times
10^{12}\,M_\odot$.

Unless otherwise noted, in this paper we set the prefactor
$\bar\sigma(M_G)$ in Eq.~(\ref{factorize}) by choosing 
$M_G=10^{12}\,M_\odot$.  Using the WMAP-5 parameters and CMBFAST, 
we find that at the present cosmic time
$\sigma(10^{12}\,M_\odot) \approx 2.03$.  This corresponds to 
$\sigma(10^{12}\,M_\odot,\tau_{1000})\approx 7.35\times 10^{-3}$.  

We are now prepared to display the results, plotting $P(\Lambda)$
as determined by Eq.~(\ref{PocketBased}). We first reproduce the
standard distribution of $\Lambda$, which corresponds to the case
when $\Lambda>0$.  This is shown in Fig.~\ref{fig:ccpocketL>0}. 
We see that the value of $\Lambda$ that we measure is between one
and two standard deviations from the mean.  Throughout the paper,
the vertical bars in the plots merely highlight the observed
value of $\Lambda$ and do not indicate its experimental
uncertainty.  The quality of the fit depends on the choice of
scale $M_G$; in particular, choosing smaller values of $M_G$
weakens the fit~\cite{Loeb,VP}.  Note however that the value of
$M_G$ that we use is already less than that recommended by
Ref.~\cite{Peacock}.

Fig.~\ref{fig:ccpocket} shows the distribution of $\Lambda$ for
positive and negative values of $\Lambda$.  We use
$\Delta\tau=5\times 10^9$ years, corresponding roughly to the age
of our solar system.  The left column corresponds to choosing
anthropic hypothesis $A$ while the right column corresponds to
anthropic hypothesis $B$.  To address the question of whether the
observed value of $|\Lambda|$ lies improbably close to the special
point $\Lambda=0$, in the second row we plot the distributions
for $P(|\Lambda|)$.  We see that the observed value of $\Lambda$
lies only a little more than one standard deviation from the
mean, which is certainly acceptable.  (Another measure of the
``typicality'' of our value of $\Lambda$ has been studied in
Ref.~\cite{VP}).

\boldmath
\subsection{Distribution of $\Lambda$ using the scale-factor cutoff}      
\label{ssec:sfcutoffL}
\unboldmath

We now turn to the calculation of $P(\Lambda)$ using a
scale-factor cutoff to regulate the diverging volume of the
multiverse.  When we restrict attention to the evolution of a
small thermalized patch, a cutoff at scale-factor time $t_c$
corresponds to a proper time cutoff $\tau_c$, which depends on
$t_c$ and the time at which the patch thermalized, $t_*$.  Here
we take the thermalized patch to be small enough that
scale-factor time $t$ is essentially constant over hypersurfaces
of constant $\tau$.  Then the various proper and scale-factor
times are related by
\bea
t_c - t_* = \int_{\tau_*}^{\tau_c}\! H(\tau)\,d\tau = \ln\big[
\tilde{a}(\tau_c)/\tilde{a}(\tau_*)\big]\,.
\label{tauc}
\eea  

Recall that all of the thermalized regions of interest share a
common evolution up to the proper time $\tau_m$, after which they
follow Eqs.~(\ref{a}).  Solving for the proper time cutoff
$\tau_c$ gives
\bea
\tau_c = \frac{2}{3}H_\Lambda^{-1}{\rm arcsinh}\!\left[
\textstyle{\frac{3}{2}}H_\Lambda\tau_m\, 
e^{\frac{3}{2}(t_c-t_*-C)} \right] ,
\label{taucL>0}
\eea  
for the case $\Lambda> 0$\,, and
\bea
\tau_c = \frac{2}{3}H_\Lambda^{-1}{\rm arcsin}\!\left[
\textstyle{\frac{3}{2}}H_\Lambda\tau_m\, 
e^{\frac{3}{2}(t_c-t_*-C)} \right] ,
\label{taucL<0}
\eea
for $\Lambda < 0$.  The term $C$ is a constant that accounts for
evolution from time $\tau_*$ to time $\tau_m$.  Note that as
$t_c-t_*$ is increased in Eq.~(\ref{taucL<0}), $\tau_c$ grows
until it reaches the time of scale-factor turnaround in the
pocket, $\tau_{\rm turn}=\pi/3H_\Lambda$, after which the
expression is ill-defined.  Physically, the failure of
Eq.~(\ref{taucL<0}) corresponds to when a thermalized region
reaches turnaround before the scale-factor time reaches its
cutoff at $t_c$.  After turnaround, the scale factor decreases;
therefore these regions evolve without a cutoff all the way up to
the time of crunch, $\tau_{\rm crunch}=2\pi/3H_\Lambda$. 

\begin{figure}[t!]
\includegraphics[width=0.4\textwidth]{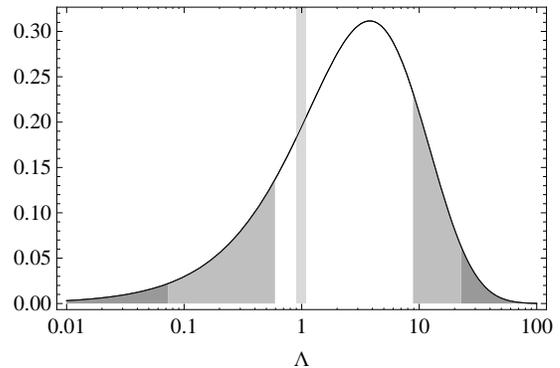}
\caption{The normalized distribution of $\Lambda$ for $\Lambda>0$, 
with $\Lambda$ in units of the observed value, for the scale-factor 
cutoff. The vertical bar highlights the value we measure, while the 
shaded regions correspond to points more than one and two standard 
deviations from the mean.}
\label{fig:ccsfcutoffL>0}
\end{figure}

When counting the number of observers in the various pockets
using a scale-factor cutoff, one must keep in mind the dependence
on the thermalized volume $V_*$ in Eq.~(\ref{Nobs1}), since in
this case $V_*$ depends on the cutoff.  As stated earlier, we
assume the rate of thermalization for pockets containing
universes like ours is independent of $\Lambda$.  Thus, the total
physical volume of all regions that thermalized between times
$t_*$ and $t_*+dt_*$ is given by Eq.~(\ref{dV}), and is
independent of $\Lambda$.  Using Eq.~(\ref{Nobs1}) to count the
number of observers in each thermalized patch, and summing over
all times below the cutoff, we find
\bea
P(\Lambda) \propto
\int_{-\infty}^{t_c}\! F\big[M_G,\tau_c(t_c,t_*)-\Delta\tau\big]\,
e^{\gamma t_*} dt_* \,. \,\,
\label{PL}
\eea
Note that regions thermalizing at a later time $t_*$ have a
greater weight $\propto e^{\gamma t_*}$.  This is an expression
of the youngness bias in the scale-factor measure.  The $\Lambda$
dependence of this distribution is implicit in $F$, which depends
on $\delta_c(\Lambda,\tau_c-\Delta\tau)/
\sigma_{\rm rms}(\Lambda,\tau_c-\Delta\tau)$, and in turn 
on $\tau_c(\Lambda)$, which is described below. 

\begin{figure*}[t!]
\begin{tabular}{ccc}
\vspace{5pt} anthropic hypothesis $A$ & & anthropic hypothesis $B$ \\
\vspace{-5pt}
\!\!\!\!\!\includegraphics[width=0.39\textwidth]{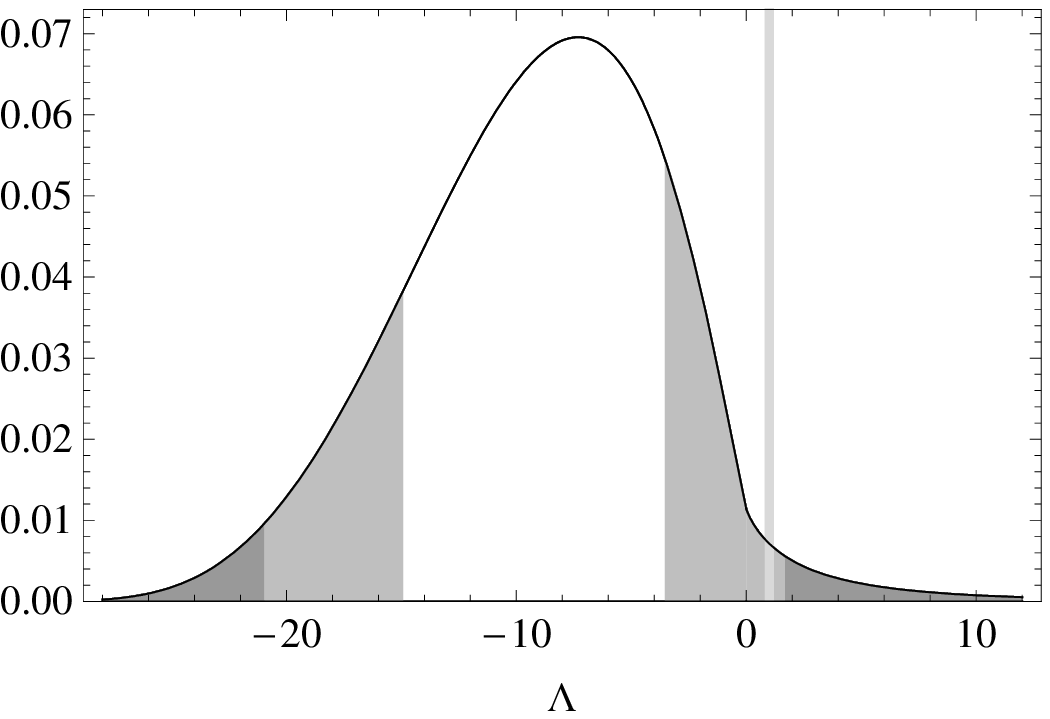} & 
\hspace{1cm} &
\includegraphics[width=0.4\textwidth]{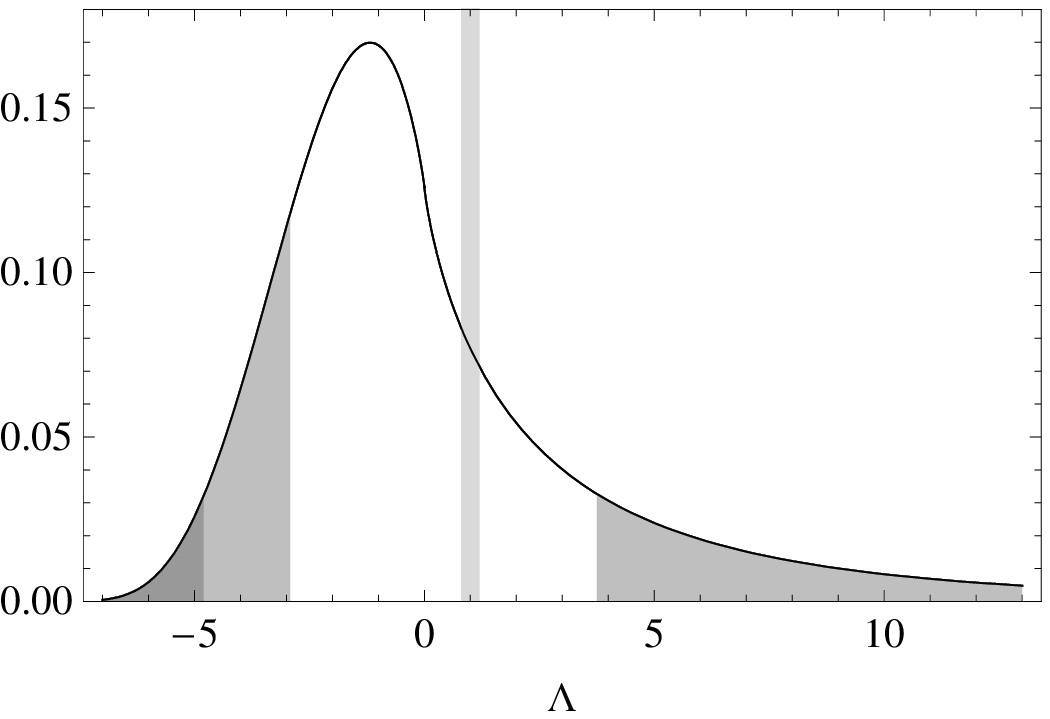} \\
\includegraphics[width=0.4\textwidth]{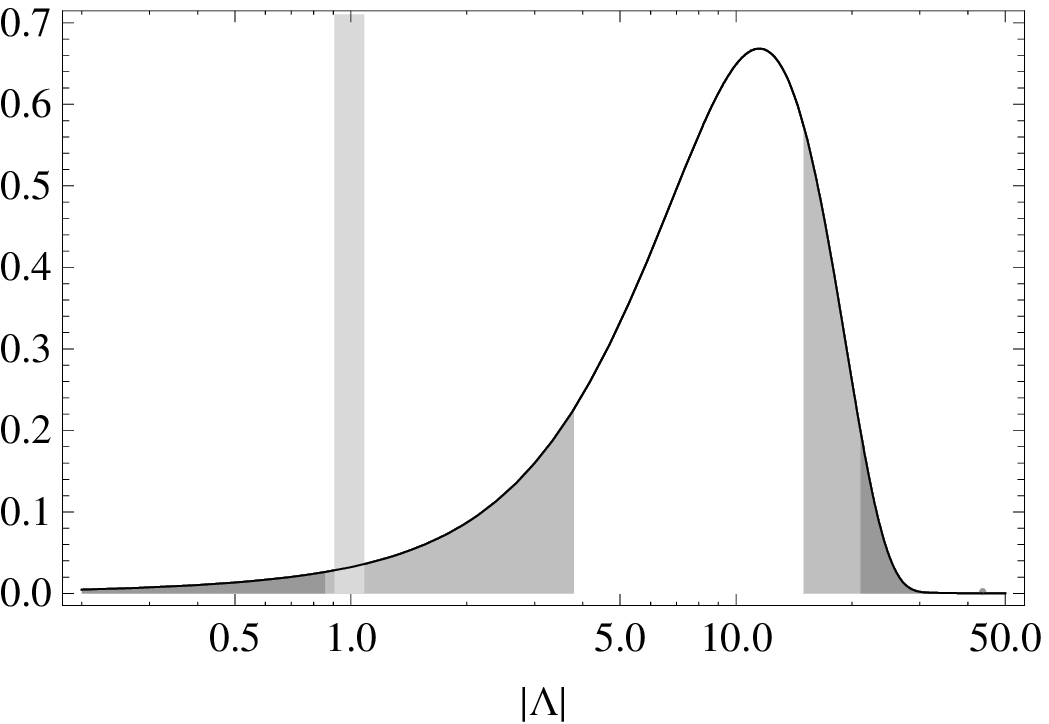} & &
\includegraphics[width=0.4\textwidth]{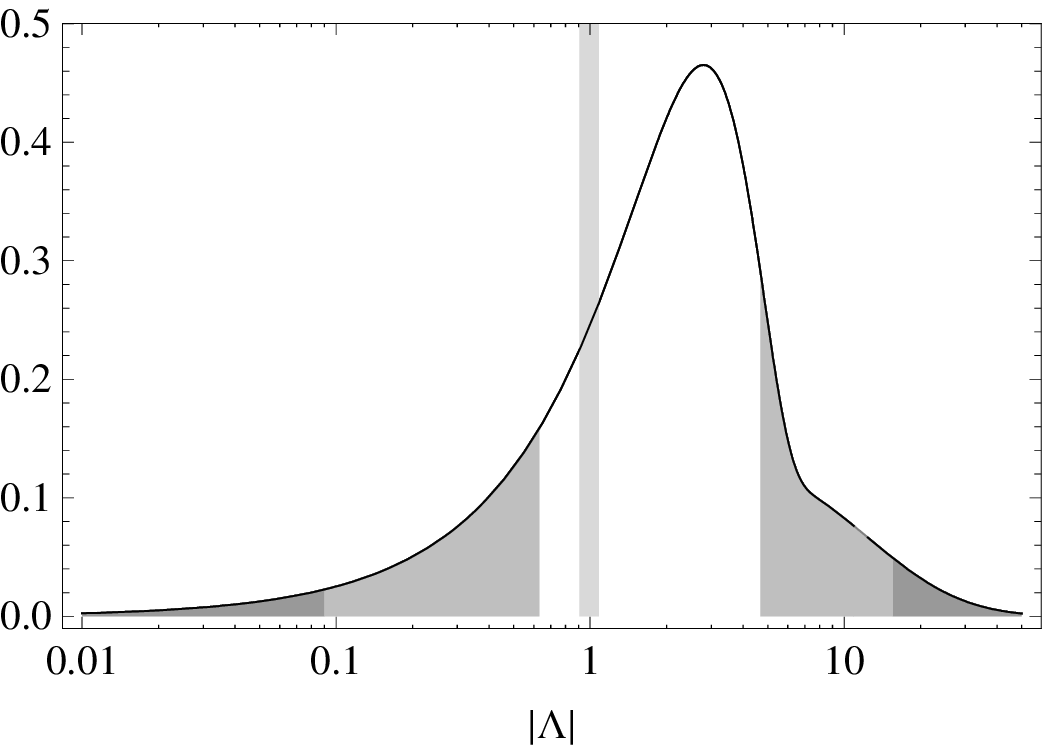} \\
\end{tabular}
\caption{The normalized distribution of $\Lambda$, with $\Lambda$ 
in units of the observed value, for the scale-factor cutoff. The
left column corresponds to anthropic hypothesis $A$ while the
right column corresponds to anthropic hypothesis $B$.  Meanwhile, 
the top row shows $P(\Lambda)$ while the bottom row shows 
$P(|\Lambda|)$. The vertical bars highlight the value we measure, 
while the shaded regions correspond to points more than one and 
two standard deviations from the mean.}
\label{fig:ccsfcutoff}
\end{figure*}

For pockets with $\Lambda> 0$, the cutoff on proper time $\tau_c$
is given by Eq.~(\ref{taucL>0}).  Meanwhile, when $\Lambda < 0$,
$\tau_c$ is given by Eq.~(\ref{taucL<0}), when that expression is
well-defined. In practice, the constant $C$ of
Eqs.~(\ref{taucL>0}) and (\ref{taucL<0}) is unimportant, since a
negligible fraction of structures form before the proper time
$\tau_m$.  Furthermore, for a reference time $\tau_m$ chosen deep
in the era of matter domination, the normalized distribution is
independent of $\tau_m$.  As mentioned above, for sufficiently
large $t_c-t_*$ Eq.~(\ref{taucL<0}) becomes ill-defined,
corresponding to the thermalized region reaching its crunch
before the scale-factor cutoff.  In this case we set
$\tau_c=\tau_{\rm crunch}$ or $\tau_c=\tau_{\rm turn}$,
corresponding to the anthropic hypothesis $A$ or $B$ described
above. 

To compare with previous work, we first display the distribution
of positive $\Lambda$ in Fig.~\ref{fig:ccsfcutoffL>0}.  We have
set $\gamma=3$ and use $\Delta\tau=5\times 10^9$ years.  Clearly,
the scale-factor cutoff provides an excellent fit to observation,
when attention is limited to $\Lambda>0$.  Note that the
scale-factor-cutoff distribution exhibits a much faster fall off
at large $\Lambda$ than the pocket-based distribution in
Fig.~\ref{fig:ccpocketL>0}.  The reason is not difficult to
understand.  For larger values of $\Lambda$, the vacuum energy
dominates earlier.  The universe then begins expanding
exponentially, and this quickly triggers the scale-factor cutoff. 
Thus, pockets with larger values of $\Lambda$ have an earlier
cutoff (in terms of the proper time) and have less time to evolve
observers.  This tendency for the cutoff to kick in soon after
$\Lambda$-domination may help to sharpen the anthropic
explanation~\cite{GLV,Bludman} of the otherwise mysterious fact
that we live so close to this very special epoch
(matter-$\Lambda$ equality) in the history of the universe.

\begin{figure*}[t!]
\begin{tabular}{ccc}
\includegraphics[width=0.4\textwidth]{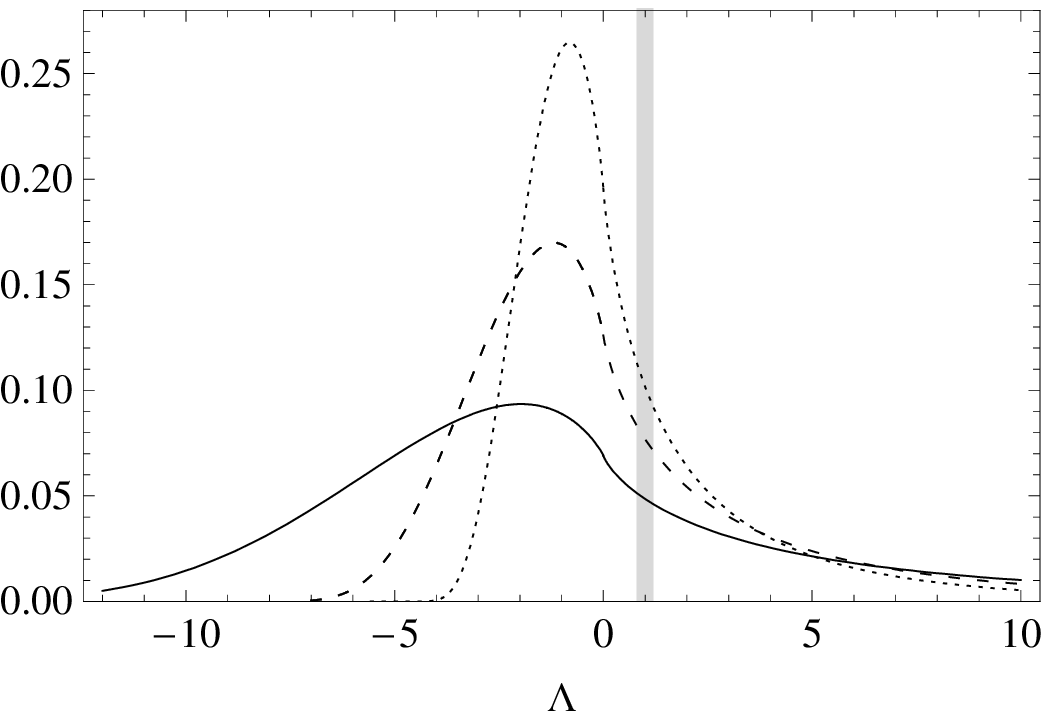} & \hspace{1cm} &
\includegraphics[width=0.4\textwidth]{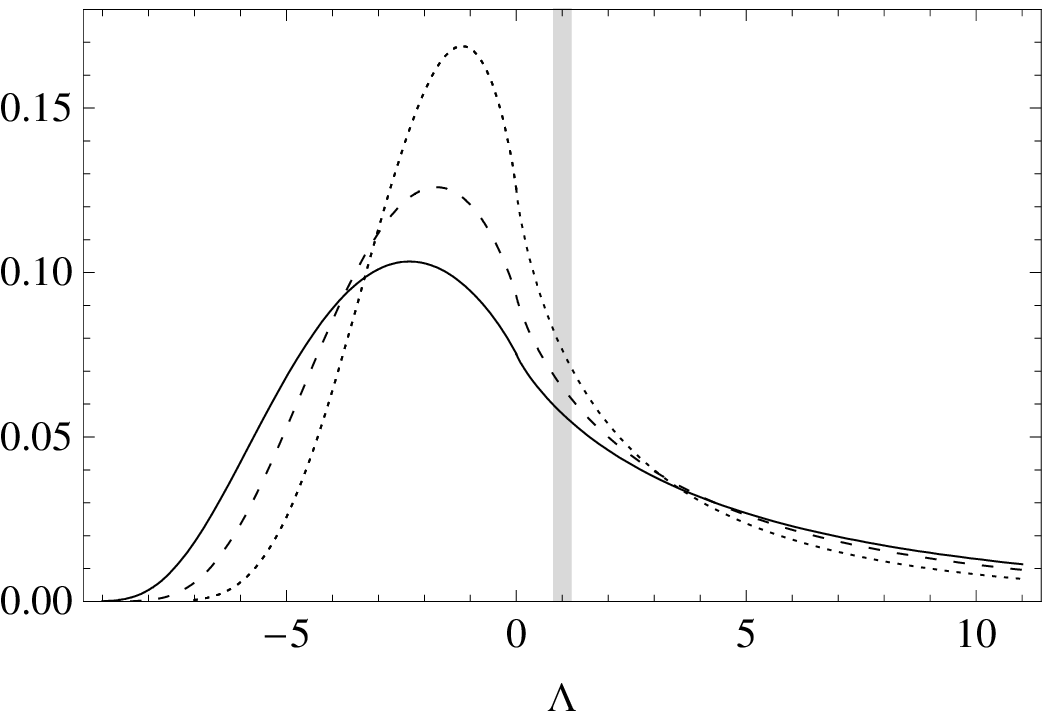} \\
\end{tabular}
\caption{The normalized distribution of $\Lambda$, with $\Lambda$ 
in units of the observed value, for anthropic hypothesis $B$ in
the scale-factor cutoff.  The left panel displays curves for
$\Delta\tau=3$ (solid), $5$ (dashed), and $7$ (dotted) $\times
10^9$ years, with $M_G= 10^{12}\,M_\odot$, while the right panel
displays curves for $M_G=10^{10}\,M_\odot$ (solid),
$10^{11}\,M_\odot$ (dashed), and $10^{12}\,M_\odot$ (dotted), with
$\Delta\tau=5\times 10^9$ years.  The vertical bars highlight
the value of $\Lambda$ that we measure.}
\label{fig:ccsfcutofftm}
\end{figure*}

The distribution of $\Lambda$ for positive and negative values of
$\Lambda$ is displayed in Fig.~\ref{fig:ccsfcutoff}, using the
same parameter values as before.  We see that the distribution
with anthropic hypothesis $A$ provides a reasonable fit to
observation, with the measured value of $\Lambda$ appearing just
within two standard deviations of the mean.  Note that the weight
of this distribution is dominated by negative values of
$\Lambda$, yet anthropic hypothesis $A$ may not give the most
accurate accounting of observers in pockets with $\Lambda<0$. 
Anthropic hypothesis $B$ provides an alternative count of the
number of observers in regions that crunch before the cutoff, and
we see that the corresponding distributions provide a very good
fit to observation.  This is the main result of this work.

The above distributions all use $\Delta\tau=5\times 10^9$ years
and $M_G=10^{12}M_\odot$.  These values are motivated
respectively by the age of our solar system and by the mass of
our galactic halo, the latter being a good match to an empirical
fit determining the halo mass scale characterizing efficient star
formation~\cite{Peacock}.  Yet, to illustrate the dependence of
our main result on $\Delta\tau$ and $M_G$, in
Fig.~\ref{fig:ccsfcutofftm} we display curves for anthropic
hypothesis $B$, using $\Delta\tau=3$, $5$, and $7\times 10^9$
years and using $M_G=10^{10}$, $10^{11}$, and $10^{12}M_\odot$. 
The distribution varies significantly as a result of these
changes, but the fit to the observed value of $\Lambda$ remains
good.

\section{Conclusions}
\label{sec:conclusions}

To date, several qualitatively distinct measures have been
proposed to regulate the diverging volume of the multiverse. 
Although theoretical analysis has not provided much guidance as
to which of these, if any, is correct, the various regulating
procedures make different predictions for the distributions of
physical observables.  Therefore, one can take an empirical
approach, comparing the predictions of various measures to our
observations, to shed light on what measures are on the right
track.  With this in mind, we have studied some aspects of a
scale-factor cutoff measure.  This measure averages over the
spacetime volume in a comoving region between some initial
spacelike hypersurface $\Sigma$ and a final hypersurface of
constant time, with time measured in units of the local Hubble
rate along the comoving geodesics.  At the end of the
calculation, the cutoff on scale-factor time is taken to
infinity.  We shall now summarize what we have learned about the
scale-factor measure and compare its properties to those of other
proposed measures.

The main focus of this paper has been on the probability 
distribution for the cosmological constant $\Lambda$.  Although 
the statistical distribution of $\Lambda$ among states in the
landscape is assumed to be flat, imposing a scale-factor cutoff
modulates this distribution to prefer smaller values of
$\Lambda$.  Combined with appropriate anthropic selection
effects, this gives a distribution of $\Lambda$ that is in a good
fit with observation.  We have calculated the distribution for
positive and negative values of $\Lambda$, as well as for the
absolute value $|\Lambda|$.  For $\Lambda >0$, we adopted the 
standard assumption that the number of observers is proportional 
to the fraction of matter clustered in halos of mass greater than 
$10^{12}M_\odot$, and allowed a fixed proper time interval 
$\Delta\tau= 5\times 10^9$ years for the evolution of observers in 
such halos.  For $\Lambda<0$, we considered two possible scenarios, 
which probably bracket the range of reasonable possibilities.  The 
first (scenario $A$) assumes that observations can be made all the 
way to the big crunch, so we count all halos formed prior to time 
$\Delta\tau$ before the crunch.  The second (scenario $B$) assumes 
that the contracting negative-$\Lambda$ phase is hazardous to life, 
so we count only halos that formed at time $\Delta\tau$ or earlier 
before the turnaround.

Our results show that the observed value of $\Lambda$ is within 
two standard deviations from the mean for scenario $A$, and within 
one standard deviation for scenario $B$.  In the latter case, the 
fit is better than that obtained in the ``standard''
calculations~\cite{Efstathiou,MSW,GLV,Tegmark,VP,Peacock}, which
assume no time cutoff (this is equivalent to choosing a
pocket-based measure on the multiverse).  The causal patch
measure also selects for smaller values of $\Lambda$ providing,
in the case of positive $\Lambda$, a fit to observation similar
to that of the scale-factor cutoff~\cite{ccbousso}.  Note,
however, that the approach of Ref.~\cite{ccbousso} used an
entropy-based anthropic weighting (as opposed to the
structure-formation-based approach used here) and that the
distribution of negative $\Lambda$ has not been studied in this
measure.

We have verified that our results are robust with respect to
changing the parameters $M_G$ and $\Delta\tau$.  The agreement
with the data remains good for $M_G$ varying between $10^{10}$
and $10^{12} M_\odot$ and for $\Delta\tau$ varying between
$3\times 10^9$ and $7\times 10^9$ years.

We have also shown that the scale-factor cutoff measure does not
suffer from some of the problems afflicting other proposed
measures.  The most severe of these is the ``youngness paradox'' 
--- the prediction of an extremely youth-dominated distribution
of observers --- which follows from the proper-time cutoff
measure.  The scale-factor cutoff measure, on the other hand,
predicts only a very mild youngness bias, which is consistent
with observation.  Another problem, which arises in pocket-based
and ``gauge-invariant'' measures, is the $Q$ catastrophe, where
one expects to measure the amplitude of the primordial density
contrast $Q$ to have an unfavorably large or small value.  This
problem ultimately stems from an exponential preference for a
large number of e-folds of slow-roll inflation in these measures. 
The scale-factor cutoff does not strongly select for more
inflation, and thus does not suffer from a $Q$ catastrophe.  An
unattractive feature of causal patch and comoving-volume measures
is that their predictions are sensitive to the assumptions one
makes about the initial conditions for the multiverse. 
Meanwhile, the scale-factor cutoff measure is essentially
independent of the initial state.  This property reflects the
attractor character of eternal inflation: the asymptotic
late-time evolution of an eternally inflating universe is
independent of the starting point.

As mentioned above, a key features of the scale-factor cutoff
measure is that, unlike the pocket-based or ``gauge-invariant''
measures, it does not reward large amounts of slow-roll
inflation.  As a result, it allows for the possibility of a
detectable negative curvature.  This issue will be discussed in
detail in Ref.~\cite{Omeganext}.

With any measure over the multiverse, one must be wary that it
does not over-predict ``Boltzmann brains'' --- observers that pop
in and out of existence as a result of rare quantum
fluctuations~\cite{BBs}.  This issue has not been addressed here,
but our preliminary analysis suggests that, with some mild 
assumptions about the landscape, the scale-factor cutoff measure 
does not have a Boltzmann brain problem.  We shall return to this 
issue in a separate publication~\cite{BBnext}.

\begin{acknowledgments}
We thank Raphael Bousso, Ben Freivogel, Andrei Linde, John
Peacock, Delia Schwartz-Perlov, Vitaly Vanchurin, and Serge 
Winitzki for useful comments and discussions.  The work of ADS is  
supported in part by the INFN ``Bruno Rossi'' Fellowship.  ADS 
and AHG are supported in part by the U.S.  Department of Energy 
under contract No. DE-FG02-05ER41360.  MPS and AV are supported in 
part by the U.S. National Science Foundation under grant NSF 322.
\end{acknowledgments}

\appendix

\boldmath
\section{Independence of the initial state}      
\label{sec:initial}
\unboldmath

In section~\ref{sec:sfcutoff} we assumed that the landscape is
irreducible, so that any vacuum is accessible through quantum
diffusion or bubble nucleation from any other (de Sitter) vacuum. 
If instead the landscape splits into several disconnected
sectors, the scale-factor cutoff can be used to find the
probability distributions $P_j^{(A)}$ in each of the sectors
(labeled by $A$).  These distributions are determined by the
dominant eigenstates of the Fokker-Planck or master equation,
which correspond to the largest eigenvalues $\gamma_A$, and are
independent of the choice of the initial hypersurfaces $\Sigma_A$
that are used in implementing the scale-factor cutoff.  But the
question still remains, how do we compare the probabilities of
vacua belonging to different sectors?

Since different sectors are inaccessible from one another, the
probability $P_A$ of being in a given sector must depend on the
initial state of the universe. For definiteness, we shall assume
here that the initial state is determined by the wave function of
the universe, although most of the following discussion should
apply to any theory of initial conditions.  According to both
tunneling
\cite{tunneling} and Hartle-Hawking \cite{HH} proposals for the wave
function, the universe starts as a 3-sphere $S_\alpha$ filled
with some positive-energy vacuum $\alpha$.  The radius of the
3-sphere is $r_\alpha=H_\alpha^{-1}$, where $H_\alpha$ is the de
Sitter expansion rate.  The corresponding nucleation probability
is
\bea
P_{\rm nucl}^{(\alpha)}\propto
\exp\left(\pm{\pi\over{H_\alpha^2}} \right)
\,,
\label{Pnucl}
\eea
where the upper sign is for the Hartle-Hawking and the lower is
for the tunneling wave function.  Once the universe has
nucleated, it immediately enters de Sitter inflationary
expansion, transitions from $\alpha$ to other vacua, and
populates the entire sector of the landscape to which the vacuum
$\alpha$ belongs.  We thus have an ensemble of eternally
inflating universes with initial conditions at 3-surfaces
$S_\alpha$ and the probability distribution $P_{\rm
nucl}^{(\alpha)}$ given by Eq.~(\ref{Pnucl}). 

If the landscape were not disconnected, we could apply the scale
factor cutoff measure to any single component $\alpha$ of the
initial wave function, and the result would be the same in all
cases.  To generalize the scale-factor cutoff measure to the
disconnected landscape, the most straightforward prescription is
to apply the scale factor cutoff directly to the initial
probability ensemble.  In that case,
\bea
P_{j, A}\propto \lim_{t_c\to\infty} \sum_{\alpha\in A} P_{\rm
nucl}^{(\alpha)}\, {\mathcal N}_j^{(\alpha)}(t_c) \,.
\label{PjA}
\eea 
Here,
\bea
{\mathcal N}_j^{(\alpha)}(t_c) = R_\alpha^{(A)} P_j^{(A)}
e^{\gamma_A t_c}
\label{Nj}
\eea
is the number of relevant observations in the entire closed
universe, starting from the hypersurface $S_\alpha$, with a
cutoff at scale-factor time $t_c$. The $R_\alpha^{(A)}$ are
determined by the initial volume of the 3-surface $S_\alpha$, and
also by the efficiency with which this initial state couples to
the leading eigenvector of Eq.~(\ref{dV}).  In other words, the
${\mathcal N}_j^{(\alpha)}(t_c)$ are calculated using $S_\alpha$
as the initial hypersurface $\Sigma_A$.  Note that only the
overall normalization of ${\mathcal N}_j^{(\alpha)}$ depends on
the initial vacuum $\alpha$; the relative probabilities of
different vacua in the sector do not.  In the limit of
$t_c\to\infty$, only the sectors corresponding to the largest of
all dominant eigenvalues,
\bea
\gamma_{\rm max}={\rm max} \{\gamma_A \} \,,
\label{gammamax}
\eea
have a nonzero probability.  If there is only one sector with
this eigenvalue, this selects the sector uniquely.

Since the issue of initial state dependence is new, one might
entertain an alternative method of dealing with the issue, in
which the probability $P_A$ for each sector is determined
immediately by the initial state, with
\bea
P_A \propto \sum_{\alpha\in A} P_{\rm nucl}^{(\alpha)}\, .
\label{psubA}
\eea
Then one could calculate any probability of interest within each
sector, using the standard scale factor cutoff method, and weight
the different sectors by $P_A$.  However, although this
prescription is well-defined, we would advocate the first method
that we described as the natural extension of the scale factor
cutoff measure.  First, it seems to be more closely related to
the description of the scale-factor cutoff measure in a connected
landscape: the only change is to replace the initial state by an
ensemble of states, determined in principle by one's theory of
the initial wave function.  Second, in a toy theory, one could
imagine approaching a disconnected landscape from a connected
one, by gradually decreasing all the cross-sector tunneling rates
to zero.  In that case, the limit clearly corresponds to the
first description, where one sector is selected uniquely if it
has the largest dominant eigenvalue.

Assuming the first of these prescriptions, the conclusion is that
the probability distribution (\ref{PjA}) defined by the
scale-factor measure is essentially independent of the initial
distribution (\ref{Pnucl}).  Some dependence on $P_{\rm
nucl}^{(\alpha)}$ survives only in a restricted class of models
where the landscape splits into a number of sectors with strictly
zero probability of transitions between them and, in addition,
where the maximum eigenvalue $\gamma_{\rm max}$ is degenerate. 
Even then, this dependence is limited to the relative probability
of the sectors characterized by the eigenvalue $\gamma_{\rm
max}$.

\boldmath
\section{The Collapse Density Threshold $\delta_c$}
\label{sec:threshold}
\unboldmath

The collapse density threshold $\delta_c$ is determined by comparing
the linearized evolution of matter perturbations with the
nonlinear evolution of a spherical top-hat density perturbation,
which can be treated as a closed FRW universe.  The collapse
density threshold $\delta_c(\tau)$ is defined as the amplitude
reached by the linear evolution of an overdensity $\delta \equiv
\delta \rho_m/\rho_m$ that has the same initial amplitude as a
top-hat density perturbation that collapses to a singularity in
proper time $\tau$. In a matter-dominated universe with zero
cosmological constant, $\delta_c$ is a constant; however, it is
well known that $\delta_c$ depends on the collapse time when
$\Lambda$ is nonzero (see e.g. Refs.~\cite{dc,PTV}).  In this
appendix we first outline the calculation of the time evolution
of $\delta_c$, then display the results for positive and negative
$\Lambda$, and finally describe how we apply it in our analysis
of the collapse fraction $F$ of Eq.~(\ref{warren}).

As suggested by the definition above, both linear and nonlinear
analyses are involved at different stages of the calculation of
the collapse density.  Arbitrarily small perturbations obey
linearized equations of motion, and their evolution defines the
linear growth function $G_\Lambda(\tau)$:
\bea
\delta(\tau) \propto G_\Lambda(\tau)\,,
\label{deltatau}
\eea
where $G_\Lambda(\tau)$ is normalized so that the behavior for
small $\tau$ is given by $G_\Lambda(\tau) \sim 
(3 H_\Lambda \tau/2)^{2/3}$, where $H_\Lambda=\sqrt{|\Lambda|/3}$.
The exact nonlinear analysis is used to determine the time at
which an overdensity with a given initial amplitude will collapse
to a singularity. For simplicity, this is worked out for the
``top-hat'' model, where the overdensity is assumed to be uniform
and spherically symmetric. Such a region is embedded in a flat
FRW universe containing only non-relativistic matter and
cosmological constant. 

By Birkhoff's theorem, the evolution of the spherical overdensity
is equivalent to that of a closed FRW universe. The Friedmann
equation for a closed FRW universe, with scale factor $a$, may be
written as:
\bea
H^2= H_\Lambda^2\left[ {\rm sign}(\Lambda) +
\frac{B(\kappa)}{a^3} - \frac{\kappa}{a^2}\right] \,,
\label{friedmann}
\eea
where $H=d\ln a/d\tau=\dot a/a$ and $B(\kappa)$ is an arbitrary
quantity that fixes the normalization of $a$.  We will always
choose $B(0)=1$, so for $\kappa=0$ the scale factor is normalized
in such a way that $\rho_m=|\rho_\Lambda|$ at $a=1$, where
$\rho_\Lambda=\Lambda/(8\pi)$ is the vacuum energy density.

Let us first focus our attention on the evolution of a linearized
density perturbation in a flat FRW universe with positive
cosmological constant; the case with negative cosmological
constant proceeds similarly. Consider a closed FRW universe
obtained by ``perturbing'' the flat universe with a small
curvature term $\delta \kappa$.  The proper time parameter
$\hat\tau$ in such a universe, as a function of the scale
factor, is given by an expansion with respect to the flat
background: $\hat\tau(a)=\tau(a)+\delta\tau(a)$, where to
linear order in $\delta \kappa$
\bea
\delta \tau(a)={\delta \kappa \over 2 H_\Lambda}\int_0^a
\frac{\sqrt{a'} \left[a' - {d B \over d \kappa} (0) \right] \,da'}
{\left(1+a'^3\right)^{3/2}}\,.
\label{deltataua}
\eea
The scale factor of the closed universe is obtained by inverting
the function $\hat\tau(a)$:
\bea
\hat a(\tau)=a(\tau)-\dot a(\tau)\, \delta\tau\bigl(a(\tau)\bigr) \,.
\label{deltaa}
\eea
As mentioned above, the evolution of this closed FRW universe
also gives the evolution of a small density perturbation.  Using
$\rho_m = (3 / 8 \pi) H_\Lambda^2 B(\kappa)/a^3$, one has
\bea
\delta={\delta \rho_m\over \rho_m}=-3{\delta a\over a}+{d B \over
d \kappa} (0) =3 H \delta \tau +{d B \over d \kappa} (0) \,, \,\,
\label{deltadeltatau}
\eea
where the last equality follows from Eq.~(\ref{deltaa}).  From 
here on, unless noted otherwise, we normalize $a$ so that 
$B(\kappa)=1$.  It is convenient to introduce the ``time'' variable
\bea
x\equiv \frac{|\rho_\Lambda|}{\rho_m}=a^3\,,
\label{defx}
\eea
for both choices of the sign of $\Lambda$.  To be consistent with
Eq.~(\ref{friedmann}), the solutions for $\kappa=0$ are not
normalized as in Eq.~(\ref{a}), but instead are given by
\bea
a(\tau)=\left\{ 
\begin{array}{l l}
\sinh^{2/3}\!
\big(\textstyle{\frac{3}{2}}H_\Lambda\tau\big)\quad 
& {\rm for }\,\, \Lambda > 0 \\
\sin^{2/3}\!
\big(\textstyle{\frac{3}{2}}H_\Lambda\tau\big)\quad
& {\rm for }\,\, \Lambda < 0 \,.
\end{array}
\right.
\label{anew}
\eea
We can then find the evolution function $\delta(x)$ from 
Eq.~(\ref{deltadeltatau}), using Eq.~(\ref{deltataua}) and also 
Eq.~(\ref{friedmann}) with $\kappa=0$:
\bea
\delta(x)&=&{1\over 2}\delta \kappa \sqrt{1+{1\over x}}\,\int_0^x
{dy\over y^{1/6}(1+y)^{3/2}}\nn\\ &=&{3\over 5}\,\delta \kappa\,
G^+(x)\,,
\label{35delta}
\eea
where the linear growth function (for $\Lambda>0$),
\bea
G^+(x)={5\over 6}\sqrt{1+{1\over x}}\,\int_0^x {dy\over
y^{1/6}(1+y)^{3/2}}\,,
\eea
is normalized so that the behavior for small $x$ is given by 
$G^+(x)\sim x^{1/ 3}=a \sim (3 H_\Lambda\tau/2)^{2/3}$.

In the $\Lambda<0$ case, the calculation proceeds along the same
steps as before and the formula (\ref{35delta}) is indeed valid
also for negative $\Lambda$, after replacing the growth function
with $G^-(x)$.  This function now has two branches $G^-_{{\rm
I}}(x)$ and $G^-_{{\rm II}}(x)$, corresponding to the expanding
and contracting phases of the universe, respectively. The first
branch of the growth function introduces no new complications,
and is found to be
\bea
G^-_{{\rm I}}(x)={5\over 6}\sqrt{{1\over x}-1}\,\int_0^x {dy\over
y^{1/6}(1-y)^{3/2}}\,. 
\eea
For the second branch, the integration is first performed over
the whole history of the universe, from $x=0$ to $x=1$ and back
to $x=0$, and then one integrates back to the value of interest
$x$.  There is a complication, however, because for this case the
denominator in Eq.~(\ref{deltataua}) is $(1-a'^3)^{3/2}$, so the
integral diverges when the upper limit is equal to 1.  The cause
of the problem is that for $\delta
\kappa \not= 0$, $a_{\rm max}$ is no longer equal to 1.  A simple
cure is to choose $B(\kappa) = 1 + \kappa$ for this case, which
ensures that $a_{\rm max}=1$ for any $\kappa$, and which
correspondingly provides an additional term in
Eq.~(\ref{deltataua}) which causes the integral to converge. 
After some manipulation of the integrals, the result can be
written as
\bea
G^-_{{\rm II}}(x) \!\!&=&\!\! {5\over 6}\sqrt{{1\over x}-1}
\left[{4\sqrt{\pi}\Gamma\left({5\over 6}\right)\over 
\Gamma\left({1\over 3}\right)}
+\int_0^x {dy\over y^{1/6}(1-y)^{3/2}}\right].\nn\\ 
\eea
The time dependence of the linear growth functions can be made
explicit by expressing $x$ as a function of $\tau$, through
Eqs.~(\ref{defx}) and (\ref{anew}).

In practice, we carry out our calculations using fitting functions 
for the growth functions, which were devised by Peacock 
\cite{Peacock}, and which are accurate to better than 0.1\%. 
These give
\begin{widetext}
\bea
G_\Lambda^+(\tau) &\simeq& 
\tanh^{2/3}\!
\big(\textstyle{\frac{3}{2}}H_\Lambda\tau\big)
\Big[1-\tanh^{1.27}\!
\big(\textstyle{\frac{3}{2}}H_\Lambda\tau\big)\Big]^{0.82} 
+ 1.437H_\Lambda^{-2/3}\left[ 1-\cosh^{-4/3}\!
\big(\textstyle{\frac{3}{2}}H_\Lambda\tau\big)\right] \label{sigp} \\
G_\Lambda^-(\tau) &\simeq& \big(\textstyle{\frac{3}{2}} 
H_\Lambda \tau\big)^{2/3}
\left[1+0.37\left(\tau/\tau_{\rm crunch}\right)^{2.18}\right]^{-1}\!
\left[1-\left(\tau/\tau_{\rm crunch}\right)^2\right]^{-1} \,,
\label{sign}
\eea
\end{widetext}
for the cases $\Lambda>0$ and $\Lambda<0$, respectively, where the 
latter fitting formula is valid for both branches.

We are now prepared to set the calculation of $\delta_c$.  Since
the universe in Eq.~(\ref{friedmann}) can be viewed as a
``perturbation'' over a flat universe with $\delta\kappa=\kappa$, the
time evolution of the overdensity is described in general by
\bea
\delta(\tau)={3\over 5}\kappa\, G_\Lambda(\tau)\,.
\label{deltakG}
\eea
The quantity $(3/5)\kappa\, a$ quantifies the size of the initial 
inhomogeneity.

In order to find the time at which the spherical overdensity
collapses, it is convenient to determine the time of turnaround
$\tau_{\rm turn}$, corresponding to when $H=0$. The time of
collapse is then given by $2\, \tau_{\rm turn}$.  The
turnaround time is obtained by integrating Eq.~(\ref{friedmann}),
choosing $B=1$:
\bea
H_\Lambda\tau_{\rm turn}(\kappa) = \int_0^{a_{\rm turn}(\kappa)}\!
\frac{\sqrt{a}\,da}{\sqrt{{\rm sign}(\Lambda) \, a^3-\kappa\,a+1}} 
\,,\,\,\,
\label{intb}
\eea
where the scale factor at turnaround $a_{\rm turn}$ corresponds
to the smallest positive solution of
\bea
{\rm sign}(\Lambda)\, a^3_{\rm turn} -\kappa\,a_{\rm turn}+1 = 0\,.
\label{bta}
\eea 
For positive $\Lambda$, the universe will collapse only if
$\kappa>\kappa_{\rm min}\equiv 3/2^{2/3}$; for negative
$\Lambda$, perturbations that collapse before the universe has
crunched have $\kappa>0$.

The numerical evaluation of the integral in Eq.~(\ref{intb}) allows
one the extract the function $\tau_{\rm turn}(\kappa)$, which can
be inverted to give $\kappa_{\rm turn}(\tau)$, expressing the
value of $\kappa$ that leads to turnaround at time $\tau$.
Finally, the collapse density threshold as a function of the time
of collapse is read from Eq.~(\ref{deltakG}):
\bea
\delta_c (\tau)={3\over 5} \kappa_{\rm turn}(\tau/2)\, 
G_\Lambda(\tau)\,.
\label{deltacgeneral}
\eea

In the limits of small and large collapse times the above
procedure can be carried out analytically to find the limiting
values of $\delta_c$. Let us consider first the large-time
regime, corresponding to small $\kappa$.  In the case
$\Lambda>0$, the smallest $\kappa$ allowed is $\kappa_{\rm min}$;
therefore
\bea
\delta_c^+(\infty)={3\over 5} \kappa_{\rm min}\, G_\Lambda^+(\infty)
\simeq 1.629\,,
\eea
where $G^+_\Lambda(\infty)=G^+(\infty)= 5 
\Gamma(2/3)\Gamma(5/6)/(3\sqrt{\pi})\simeq 1.437$. The case 
$\Lambda<0$ is a little more complicated. The
collapse time cannot exceed $\tau_{\rm crunch}=2\pi/3 H_\Lambda$,
corresponding to $\kappa=0$. At small $\kappa$, the integral in
Eq.~(\ref{intb}) is expanded to give
\bea
H_\Lambda \tau_{\rm turn}(\kappa)\simeq \frac{1}{2}
H_\Lambda \tau_{\rm crunch}-{2\over 5} {\sqrt{\pi} 
\Gamma\left({11\over 6}\right)\over 
\Gamma\left({1\over 3}\right)}\,\kappa\,.
\label{turnksmall}
\eea
On the other hand, the growth function $G^-(\tau)$ in the
neighborhood of $\tau_{\rm crunch}$ behaves as
\bea
G^-_\Lambda(\tau\approx\tau_{\rm crunch})\simeq 
{10\over 3} {\Gamma\left({5\over 6}\right)\over
\sqrt{\pi}\Gamma\left({1\over 3}\right)}{1\over  
(1-\tau/\tau_{\rm crunch})}\,.\,\,\,
\label{Gntcrunch}
\eea
After using Eqs.~(\ref{turnksmall}) and (\ref{Gntcrunch}) in the
general formula (\ref{deltacgeneral}), we simply get
\bea
\delta_c^-(\tau_{\rm crunch})=2\,.
\eea

\begin{figure}[t!]
\includegraphics[width=0.4\textwidth]{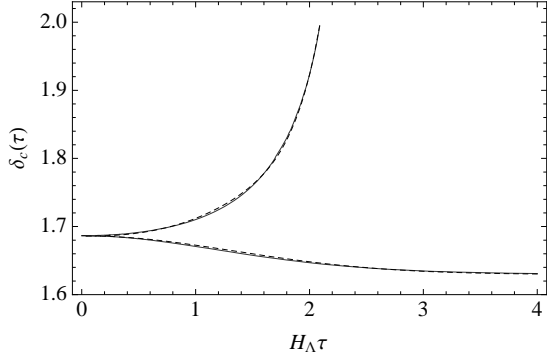}
\caption{The collapse density thresholds $\delta^+_c$ (for $\Lambda> 0$) 
and $\delta^-_c$ (for $\Lambda<0$), as functions of time.  The
solid curves represent numerical evaluations of $\delta^\pm_c$,
while the dashed curves correspond to the fitting functions in
Eq.~(\ref{fitdeltac}).  Note that $\delta^+_c$ decreases with time,
while $\delta^-_c$ increases with time.}
\label{fig:threshold}
\end{figure}

In the opposite regime $H_\Lambda \tau\ll 1$, corresponding to
large $\kappa$, the growth functions are $G^\pm_\Lambda(\tau)\simeq
a(\tau)\simeq (3H_\Lambda\tau/2)^{2/3}$. The integral
(\ref{intb}) can be analytically solved in this limit: $H_\Lambda
\tau_{\rm turn}(\kappa)=\pi/(2 \kappa^{3/2})$. Combining these
results leads to
\bea
\delta_c^\pm(0)\,=\,{3\over 5}\left({3\pi\over 2}\right)^{2/3}\!
\simeq\, 1.686\,,
\eea
which is also the constant value of $\delta_c$ in a $\Lambda=0$
universe.

The time dependence of $\delta_c$ is displayed in
Fig.~\ref{fig:threshold}, for both positive and negative values
of $\Lambda$.  We also display the following simple fitting
functions,
\bea
\delta^+_c(\tau) \!&=&\! 1.629+0.057\,e^{-0.28 H_\Lambda^2\tau^2} \nn\\
\delta^-_c(\tau) \!&=&\!
1.686 + 0.165\left({\tau\over \tau_{\rm crunch}}\right)^{\!2.5}
\!\!+0.149\left({\tau\over \tau_{\rm crunch}}\right)^{\!11} \nn\\ 
\label{fitdeltac}
\eea
which are accurate to better than 0.2\%.  Although we choose to
include the effect of the time evolution of $\delta_c$, our
results are not significantly changed by treating $\delta_c$ as a
constant.  This is easy to understand. First of all, $\delta^+_c$
varies by only about 3\%.  The evolution of $\delta^-_c$ is more
significant, about $15$\%, and most of this happens at very late
times.  But our anthropic weight in Eq.~(\ref{Nobs1}) never
samples $\delta^-_c$ within a time $\Delta\tau$ of $\tau_{\rm
crunch}$.

Finally, we point out that the appearance of $G_\Lambda(\tau)$ in
this discussion is not needed for the calculation, and appears here 
primarily to make contact with other work.  From 
Eq.~(\ref{warren}) one sees that the collapse fraction depends only 
on the ratio of $\delta_c(\tau)/\sigma(M_G,\tau)$, which from 
Eqs.~(\ref{factorize}) and~(\ref{deltacgeneral}) can be seen to equal 
$(3/5) \kappa_{\rm turn}(\tau/2)/\bar \sigma(M_G)$.  Expressed in 
this way, Eq.~(\ref{warren}) becomes fairly transparent.  Since 
$\kappa$ is a measure of the amplitude of an initial perturbation, 
Eq.~(\ref{warren}) is saying that the collapse fraction at time $\tau$ 
depends precisely on the magnitude required for an initial top-hat 
perturbation to collapse by time $\tau$.  In more detail, 
Eq.~(\ref{warren}) is predicated on a Gaussian distribution of initial 
fluctuations, where the complementary error function erfc$(x)$ is the 
integral of a Gaussian.  The collapsed fraction at time $\tau$ is given 
by the probability, in this Gaussian approximation, for the initial 
fluctuations to exceed the magnitude needed for collapse at time $\tau$.  
From a practical point of view, the use of $G_\Lambda(\tau)$ in the 
discussion of the collapse fraction can be a helpful simplification if 
one uses the approximation that $\delta_c\approx$ const.  We have not 
used this approximation, but as described above, our results would not 
be much different if we had.  We have maintained the discussion in 
terms of $G_\Lambda(\tau)$ to clarify the relationship between our 
work and this approximation.

\end{document}